\documentclass[conference,compsoc]{IEEEtran}
\newif\ifSPACEHACK
\SPACEHACKtrue 
\newif\ifDEBUG
\DEBUGfalse

\newif\ifANONYMOUS
\ANONYMOUSfalse

\usepackage[compress]{cite} % numbered ordering of multiple citations, set to compress ranges
\usepackage[dvipsnames]{xcolor}
\usepackage{amsmath,amsfonts}
\usepackage{algorithmic}
\usepackage{graphicx}
\usepackage{textcomp}
\usepackage{booktabs} % For formal tables \toprule
\usepackage{xspace} % Put an \xspace within numeric macros
\usepackage[normalem]{ulem}
\usepackage{makecell}
\usepackage{tcolorbox}
\usepackage{enumitem}
\usepackage{siunitx}
\usepackage[vskip=1em,font=itshape,leftmargin=2em,rightmargin=2em]{quoting}
\usepackage{svg}
\usepackage{listings}
\usepackage{subfigure}
\usepackage{url}
\usepackage{pifont}% http://ctan.org/pkg/pifont
\newcommand{\cmark}{{\color{ForestGreen}\ding{51}}}%
\newcommand{\xmark}{{\color{Maroon}\ding{55}}}%

\usepackage{soul}
\usepackage{hyperref}
\usepackage{cleveref}

\ifDEBUG
    \newcommand{\JD}[1]{\textcolor{blue}{[Jamie: #1]}}
    \newcommand{\Kelechi}[1]{\textcolor{olive}{[Kelechi: #1]}}
    \newcommand{\TRS}[1]{\textcolor{orange}{[Taylor: #1]}}
    \newcommand{\AK}[1]{\textcolor{purple}{[Andy: #1]}}
    \newcommand{\ST}[1]{\textcolor{red}{[Santiago: #1]}}
    \newcommand{\SB}[1]{\textcolor{red}{[Saurabh: #1]}}
    \newcommand{\Eman}[1]{\textcolor{teal}{[Eman: #1]}}
    \newcommand{\TODO}[1]{\hl{#1}}
    \newcommand{\CHOP}[1]{}

    \hypersetup{
        pdfpagemode=UseNone % don't show thumbnails or bookmarks on opening
    }

\else
    \newcommand{\JD}[1]{}
    \newcommand{\Kelechi}[1]{}
    \newcommand{\TRS}[1]{}
    \newcommand{\AK}[1]{}
    \newcommand{\ST}[1]{}
    \newcommand{\SB}[1]{}
    \newcommand{\Eman}[1]{}
    \newcommand{\TODO}[1]{}
    \newcommand{\CHOP}[1]{}
    
\fi

\newcommand{\myparagraph}[1]{\vspace{0.08cm}\noindent\hspace{0.1cm} \textit{#1:}}
\usepackage{balance} % Put \balance on the last page
\usepackage{amsmath}
\usepackage{cleveref}

\crefformat{section}{\S#2#1#3}
\crefrangeformat{section}{\S#3#1#4--\S#5#2#6}
\crefname{figure}{Figure}{Figures}
\crefname{appendix}{Appendix}{Appendices}
\crefname{table}{Table}{Tables}
\crefname{algorithm}{Algorithm}{Algorithms}
\crefname{listing}{Listing}{Listings}
\crefname{theorem}{Theorem}{Theorems}
\crefname{thm}{Theorem}{Theorems}
\crefname{lemma}{Lemma}{Lemmata}
\crefname{equation}{Eqt.}{Eqts.}
\crefformat{Grammar}{Grammar #1}

\usepackage{enumitem}
\usepackage{booktabs}

\newcommand{\ie}{\textit{i.e.,} }
\newcommand{\eg}{\textit{e.g.,} }
\newcommand{\etal}{\textit{et al.}\xspace}
\newcommand{\etals}{\textit{et al.'s}\xspace}

\usepackage{enumitem}

\DeclareUnicodeCharacter{2060}{\nolinebreak}

\newcommand{\maven}{Maven Central\xspace}
\newcommand{\mavenPackages}{499,588\xspace}
\newcommand{\mavenFilter}{243,191\xspace}
\newcommand{\mavenMeasure}{49,959\xspace}

\newcommand{\hf}{{Hugging Face}\xspace}
\newcommand{\hfPackages}{559,517\xspace}
\newcommand{\hfFilter}{128,338\xspace}
\newcommand{\hfMeasure}{128,338\xspace}
\newcommand{\hfGated}{5,138\xspace}

\newcommand{\docker}{{Docker Hub}\xspace}

\newcommand{\dockerCosignAdded}{28 Jul 2021\xspace}
\newcommand{\dockerPackages}{1,001,771\xspace}
\newcommand{\dockerFilter}{91,719\xspace}
\newcommand{\dockerMeasure}{91,719\xspace}

\newcommand{\pypi}{{PyPI}\xspace}
\newcommand{\pypiPGPRemoved}{{23 May 2023}\xspace}
\newcommand{\pypiPGPDeemp}{{22 Mar 2018}\xspace}
\newcommand{\pypiPackages}{623,346\xspace}
\newcommand{\pypiFilter}{205,513\xspace}
\newcommand{\pypiMeasure}{205,513\xspace}

\newcommand{\npm}{{NPM}\xspace}

\newcommand{\registries}{{\pypi, \maven, \docker, and \hf}\xspace}
\newcommand{\numRegistries}{{4}\xspace}

\newcommand{\startDate}{{01 Jan 2015}\xspace}
\newcommand{\finishDate}{{31 Dec 2023}\xspace}
\newcommand{\startYear}{{01 Jan 2023}\xspace}
\newcommand{\ecoDate}{{01 Mar 2024}\xspace}

\input{data/summary.tex}
\input{data/summaryyr.tex}
\input{data/crypto.tex}
\input{data/exp_keys.tex}
\input{data/metric}
\input{data/stats.tex}

\ifCLASSOPTIONcompsoc
  \usepackage[compress]{cite}
\else
  \usepackage[compress]{cite}
\fi
\ifCLASSINFOpdf
\else
\fi

\begin{document}
% \title{Artifact Signing in Five Public Software Package Registries: \\ Adoption Measurements and Influencing Factors}
% \title{Signing in Five Public Software Package Registries: \\ Adoption Measurements and Influencing Factors}
\title{Signing in Four Public Software Package Registries: \\ Quantity, Quality, and Influencing Factors}
%\title{Measuring Artifact Signing in Four Public Software Package Registries: \\ Quantity, Quality, and Influencing Factors}

%\author{
% {Anonymous.}
%{\rm Taylor R.\ Schorlemmer}\\
%Purdue University
%\and
%} % end author

\author{
	\IEEEauthorblockN{Taylor R. Schorlemmer}
	\IEEEauthorblockA{Purdue University\\
		tschorle@purdue.edu}\\
	\IEEEauthorblockN{Eman Abu Ishgair}
	\IEEEauthorblockA{Purdue University\\
		eabuishg@purdue.edu}
	\and
	\IEEEauthorblockN{Kelechi G. Kalu}
	\IEEEauthorblockA{Purdue University\\
		kalu@purdue.edu}\\
	\IEEEauthorblockN{Saurabh Bagchi}
	\IEEEauthorblockA{Purdue University\\
		sbagchi@purdue.edu}
	\and
	\IEEEauthorblockN{Luke Chigges}
	\IEEEauthorblockA{Purdue University\\
		lchigges@purdue.edu}\\
	\IEEEauthorblockN{Santiago Torres-Arias}
	\IEEEauthorblockA{Purdue University\\
		santiagotorres@purdue.edu}
	\and
	\IEEEauthorblockN{Kyung Myung Ko}
	\IEEEauthorblockA{Purdue University\\
		ko112@purdue.edu}\\
	\IEEEauthorblockN{James C. Davis}
	\IEEEauthorblockA{Purdue University\\
		davisjam@purdue.edu}
}


% compsoc conferences compared to traditional conferences), use this
%Telephone: (800) 555--1212, Fax: (888) 555--1212}


\maketitle

\begin{abstract}
	%Over the past two decades, software applications have become increasingly dependent on third-party packages distributed by third-party package registries.
	Many software applications incorporate open-source third-party packages distributed by public package registries.
	Guaranteeing authorship along this supply chain is a challenge.
	%Once an engineer identifies a suitable package, can they ensure they are actually receiving the package produced by its maintainer? % which packages can be trusted?
	Package maintainers can guarantee package authorship through \emph{software signing}.
	However, it is unclear how common this practice is, and whether the resulting signatures are created properly.
	Prior work has provided raw data on registry signing practices, but
	only measured single platforms,
	did not consider quality,
	did not consider time,
	and
	did not assess factors that may influence signing.
	%Prior work has suggested that the creation of signatures is challenging, and have performed single-platform measurements to confirm this or have proposed systems to streamline the signing process.
	% We lack a comprehensive, multi-platform understanding of signing adoption and relevant factors. % that influence it.
	We do not have up-to-date measurements of signing practices nor do we know the quality of existing signatures.
	Furthermore, we lack a comprehensive understanding of factors that influence signing adoption.
	% \JD{MUSTFIX: Needs spice here --- motivation is wimpy compared to what Intro does}
	% A software dependency involves two forms of trust --- that the software does what it claims (an undecidable problem), and that the author is who they claim.
	% This latter form of trust can be addressed through public-key cryptography, via a process called software signing.

	This study addresses this gap.
	We provide measurements across three kinds of package registries:
	traditional software (Maven, PyPI),
	container images (DockerHub),
	and
	machine learning models (Hugging Face).
	For each registry, we describe the nature of the signed artifacts as well as the current quantity and quality of signatures.
	Then, we examine longitudinal trends in signing practices.
	Finally, we use a quasi-experiment to estimate the effect that various factors had on software signing practices.
	To summarize our findings:
	(1) mandating signature adoption improves the \textit{quantity} of signatures;
	(2) providing dedicated tooling improves the \textit{quality} of signing;
	(3) getting started is the hard part --- once a maintainer begins to sign, they tend to continue doing so;
	and
	(4) although many supply chain attacks are mitigable via signing, signing adoption is primarily affected by registry policy rather than by public knowledge of attacks, new engineering standards, etc.
	These findings %shift the focus on software signing adoption away from the difficulties faced by individuals,
	highlight the importance of software package registry managers and signing infrastructure. % in signing adoption.

\end{abstract}


% \fi
\IEEEpeerreviewmaketitle

\section{Introduction}

%%%
% OSS matters and supply chains are scary
%%%
% \JD{Review Saurabh's comments on the abstract and intro --- see annotated PDF}
Commercial and government software products incorporate open-source software packages~\cite{gsa_18f_18f_nodate,department_of_defense_open_2003}. %, including in safety-critical systems~\cite{department_of_defense_open_2003}.
In a 2023 study of 1,703 commercial codebases across 17 sectors of industry, Synopsys found that 96\% used open-source code, and 76\% of the total application code was open-source~\cite{synopsys_2023_2023}.
%OSS is typically integrated into larger applications as a dependency, \eg providing utility functionality, a component, or a framework for business logic.
% Many software products include over 80\% of open-source components~\cite{ZZZ}.
%Many modern software products depend heavily on open-source components.
Open-source software packages depend on other packages, creating \emph{software supply chains}~\cite{okafor_sok}. % --- and compromises in dependencies impact product security.
Malicious actors have begun to attack software supply chains, injecting malicious code into packages to gain access to downstream systems~\cite{okafor_sok}.
These attacks have affected critical infrastructure and national security~\cite{solarwinds, solarwinds-microsoft, solarwinds-zdnet, sonatype_state_2022}.
%This is primarily because attackers, after compromising a single upstream component, gain access to several downstream systems.
%The compromise of the SolarWinds Orion supply chain and the subsequent impact on their customers was an awakening example of this fact.
%The nearly universal dependence on open-source code has further increased the viability of this attack style.
%Between 2019 and 2022, software supply chain attacks increased an average 742\% year over year~\cite{sonatype_state_2022}.

%%%
% Techniques exist to secure supply chains --- signing is one of them
%%%
Many mitigations have been proposed for software supply chain attacks.
Some approaches seek to increase confidence in a package's \textit{behavior}, \eg
measuring use of best practices~\cite{zahan_openssf_2023,zahan2023softwarePractices},
independent validation~\cite{serebryany_oss-fuzz_2017},
and
formal guarantees~\cite{10.1145/3548606.3559373};
Other approaches target the package's \textit{provenance}, \eg
Software Bill of Materials (SBOMs)~\cite{balliu2023challenges,zahan2023software}
and
``vendoring'' trusted copies of dependencies~\cite{winters_software_2020}.
The strongest guarantee of a package's provenance is a cryptographic signature by its maintainer.
Prior work has noted that many packages are unsigned~\cite{kuppusamy_diplomat, woodruff_pgp_2023}.
%, characterizing software signing practices
%These concerns have largely been due to the poor signature adoption noted by members of the open-source community.
However,
we lack up-to-date measurements of software signing practices,
and
we do not know the general quality of existing signatures.
Furthermore, we lack a deeper understanding of factors that affect adoption rates.
This knowledge would guide future efforts to incentivize software signing, so that the provenance of software supply chains can be improved.

Our work provides this knowledge: we measure software signing practices in four public software package registries, and we use that data to infer factors that influence software signing.
%We formulate a simple two-factor signing adoption theory to explain differences between registries:
%when signing is \textit{difficult}, adoption goes down;
%and
%when signing is \textit{incentivized}, signing goes up.
We selected four registries for a quasi-experiment~\cite{felderer_contemporary_2020}:\footnote{A quasi-experiment seeks cause-and-effect relationships between independent and dependent variables without subject randomization.}
two with signing policies (Maven-positive, PyPI-negative),
one with dedicated tooling (DockerHub),
and
one with no stance on signing (Hugging Face).
Under the assumption that maintainers behave similarly across registries, comparing signing practices in these registries will shed light on the factors that influence software signing.
\iffalse
	In these registries,
	we provide updated measurements of software signing quantity and quality
	and
	compare behaviors by variables.
\fi
In addition to registry-dependent variables, we consider three registry-independent factors:
organizational policy,
dedicated signing tools,
signing-related events such as high-profile cyberattacks,
and
the startup effort of signing.
%Then we test our theory through a quasi-experiment~\cite{felderer_contemporary_2020} on \numRegistries registries between \startDate and \finishDate.
%Our choice for quasi-experiment is necessitated by the fact that

% Our work --- results
Here are the highlights of our results.
Registry-specific signing policies have a large effect on signing frequency:
requiring signing yields near-perfect signing rates (Maven),
while
decreasing its emphasis reduces signing (PyPI).
%Signing quantity increases when mandated (Maven), and decreases when platforms disincentivize signatures (PyPI).
Signing remains difficult ---
only the registry with dedicated signing tools had perfect signature quality (DockerHub),
while the other three had signature quality rates of
\summary[maven][good_p] (Maven),
\summary[pypi][good_p] (PyPI),
and
\summary[huggingface][good_p] (Hugging Face).
We observed no effects from signing-related news, such as
high-profile cyberattacks
and
new engineering standards that recommend software signing.
Finally, the first signature is the hardest: after a maintainer first signs a package, they are likely to continue signing that package.
%We measured the adoption of Software signing across five package registries, including Maven Central, PyPI, NPM,
%Hugging Face, and Docker Hub, and found three interesting things
% We measured XXX from YYY and found NUMBER interesting things.
%First, Signing is relatively low in Hugging Face (Machine learning) registry compared to traditional software package registries.
%Second, Difficulty or ease of signing has little to no impact don't the adoption of signing in a package registry.
%Third,  platforms with greater incentives to sign usually see greater adoption rates by maintainers.

To summarize our contributions:
\begin{enumerate}
	\item We present up-to-date measurements of software signing practices --- quantity and quality --- in four major software package registries.
	\item We use a quasi-experiment to estimate the effect of several factors on software signing practices. Registry policies correlates with quantity. Dedicated tooling correlates with quality. Signing events do not correlate with signing practices. Starting to sign correlates with continued signing.
\end{enumerate}

%\JD{On the contributions. I think it my not be a contribution in and by itself, but something that stands out from this pper vs others is that we are doing a time-series analysis to find incentives, rather than just checking wht number the scorecards website spit out...said differently, I wonder if something like "we identify correlatitions between incentives/interventions/practices and package sigining" then "we show/confirm these correlations across different registries through a time-based series analysis" or so}

%\noindent Our data is available via an artifact.

\section{Background} \label{sec:Background}

\cref{sec:SWSupplyChains} discusses software supply chains.
\cref{sec:Signing} describes software signing, generally and in our target registries.

\subsection{Software Supply Chains} \label{sec:SWSupplyChains}
%\subsection{Software Supply Chains}

%\JD{Consider a tiny figure (steal Wenxin's) for brevity.}

%%%
% Terms
%%%
In modern software development, engineers commonly integrate and compose existing units of functionality to create novel applications~\cite{synopsys_2023_2023}.
Each unit of functionality is commonly distributed in the form of a \textit{software package}~\cite{Jiang2023PTMReuse}:
software in source code or binary representation,
accompanied by documentation,
shared under a license,
and
distinguished by a version number.
These units of functionality may be available directly from version control platforms (\eg source code~\cite{sourceforge_compare_nodate,gitlab_devsecops_nodate} or lightweight GitHub Packages~\cite{GitHubPackages})), but are more commonly distributed through separate \textit{software package registries}~\cite{jiang2022PTMSupplyChain,zimmermann2019small}.
These registries serve both
package maintainers (\eg providing storage and advertising)
and
package users (\eg indexing packages for search, and facilitating dependency management).
These facilities for software reuse result in webs of dependencies comprising the \textit{software supply chain}~\cite{enck2022top,okafor_sok_2022}.

\iffalse
	applications depend on other applications,
	and
	packages depend on other packages,
	resulting in a web of
\fi

%%%
% Motivational adoption measurements
%%%

Many empirical studies report the widespread use of software packages and the complexity of the resulting supply chains.
%This trend is well-documented in both commercial and open-source contexts.
Synopsys's 2023 Open Source Security and Risk Analysis (OSSRA) Report examined 1,703 commercial codebases across 17 industries~\cite{synopsys_2023_2023}, revealing that 96\% of these codebases incorporate third-party open-source software components, averaging 595 distinct open-source dependencies per project.
Similarly,
Kumar \etal reported that over 90\% of the top one million Alexa-ranked websites rely on external dependencies~\cite{kumar2017security},
and
Wang \etal found that 90\% of highly popular Java projects on GitHub use third-party packages~\cite{Wang_Chen_Huang_Shi_Xu_Peng_Wu_Liu_2020}.
Selecting and managing software dependencies is thus an important software engineering practice~\cite{de2008empirical, pashchenko2020qualitative}.
Software engineers must decide which packages to use in their projects, \ie what to include in their application's software supply chain~\cite{jadhav2009evaluating, jadhav2011framework}.
Engineers consider many aspects, encompassing functionality, robustness, maintainability, compatibility, popularity, and security~\cite{decan2019empirical, ghofrani2022trust,Jiang2023PTMReuse, vu_lastpymile_2021}.
Specific to security, various tools and methodologies have been proposed.
% each tailored to address the Transparency, validity, and Separation properties. 
These include
in-toto~\cite{intoto},
reproducible builds~\cite{lamb_reproducible_2022},
testing~\cite{hejderup_use_2022,benedetti_automatic_2022},
LastPyMile~\cite{vu_lastpymile_2021},
SBOMs~\cite{noauthor_minimum_nodate},
and
BuildWatch~\cite{ohm_towards_2020}.
%Okafor \etal have presented a comprehensive taxonomy of these tools and methodologies, aligning each with the respective security properties they are tailored to address --- Transparency, validity, and Separation~\cite{okafor_sok_2022}.
Okafor \etal summarized these approaches in terms of three security properties for a project's software supply chain:
validity (packages are what they claim to be),
transparency (seeing the full chain),
and
separation of concerns~\cite{okafor_sok_2022}.
Validity is a prerequisite property --- if a individual package is invalid, transparency and separation will be of limited use.

\subsection{Promoting Validity via Software Signing} \label{sec:Signing}

%Software signing is a common defensive technique against supply chain attacks.

%Software signing is a common method for establishing the provenance of artifacts in package registries.
\textit{Software signing} is the standard method for establishing the validity of packages.
Signing uses public key cryptography to bind an identity (\eg a package maintainer's private key) to an artifact (\eg a version of a package)~\cite{shirey2007internet}.
%This is achieved by using a private-public keypair~\cite{Internet-Security-Glossary} and cryptographically binding an identity (\ie a public key belonging to the package's maintainer), to a particular instance of an artifact (\eg an installable package).
With an artifact, a signature, and a public key, one can verify whether the artifact was indeed produced by the maintainer.
Software signing is a development practice recommended by industry~\cite{zahan_openssf_2023,security_technical_advisory_group_software_2021,slsa} and government~\cite{nist-2018-codesigning,cisa-2021-supplychain} leaders.

\subsubsection{Signing Process and Failure Modes} \label{sec:signing_process}

\begin{figure*}
	\centering
	\includegraphics[width=.9\linewidth]{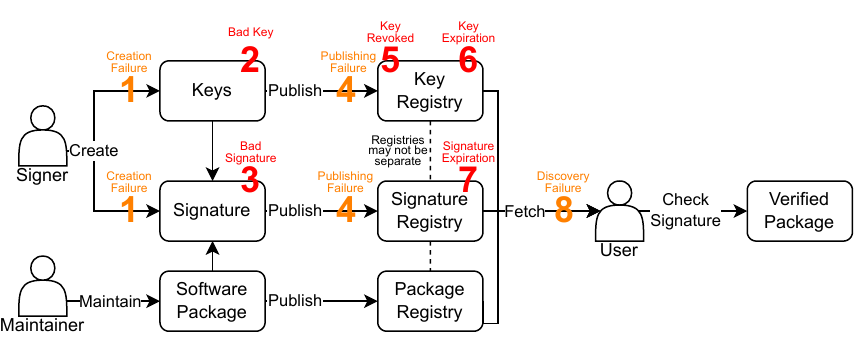}
	\caption{
		Maintainers create software packages and signers create keys which are used to create a signature.
		Each of these artifacts are published to a registry.
		Depending on ecosystem, the registries and the actors may or may not be separate.
		Users fetch these artifacts and can check signatures using infrastructure-specific tooling.
		This creates a verified package.
		Red and orange numbers indicate the failure modes described in~\cref{sec:signing_process}.
		%Signers can fail to create keys or signatures (1).
		%Keys may use an unsupported algorithm or be corrupted (2).
		%Similarly, signatures may use unsupported algorithms or do not verify a package (3).
		%The signer may neglect to publish the public key or signature (4) to a registry.
		%Public keys can be revoked (5) or expire (6).
		%Signatures can also expire (7).
		%Finally, users may be unable to discover signatures or public keys that have been published (8).
		% SB (10/15/23): "do not verify a package" I interpret as the signatures do not match indicating that the software package is not valid. This is not a failure success, it is a good case. 
		Red numbers indicate discernible failures.
		The orange numbers (modes 1, 4, and 8) are not distinguishable from one another by an external audit --- when keys are missing, we cannot determine whether they were never created (mode 1), were not published (mode 4), or were undiscoverable by us (mode 8). % look the same as signatures that have not been published (mode 4) or are undiscoverable (mode 8).
	}
	\label{fig:good_signing}
\end{figure*}

% Signatures take several steps to create and validate. These steps can be grouped into, the Signature creation, package publication, and signature validation by users.  
% \cref{fig:good_signing} illustrates in generality the signing process.  
Most software package registries require similar signing processes.
\cref{fig:good_signing} illustrates this process, beginning with a maintainer and a (possibly separate) signer.
They publish a signed package and separately the associated cryptographic material, so that a user can assess the validity of the result.

In package registries, there are two typical identities of the signer.
In the \textit{maintainer-signer} approach, the maintainer is also the signer (\eg Maven).
In the \textit{registry-signer} approach, the maintainer publishes a package and the registry signs it (\eg NPM).
These approaches trade usability against security.
Managing signatures is harder for maintainers, but the maintainer-signer approach gives the user a stronger guarantee: the user can verify they have the same package signed by the maintainer.
The registry-signer approach is easier for maintainers, but users cannot detect malicious changes made during the package's handling by the package registry.

\cref{fig:good_signing} also depicts failure modes of the signing process.
%In an effort to shed light on the complications that arise in signing software artifacts, we identify a collection of possible ways in which this process can fail and produce a \emph{low quality signature}.
%As depicted in ~\cref{fig:good_signing}, various issues can potentially arise in this process.
These modes stem from several factors, including
the complexity of the signing process,
the (non-)user-friendliness of the signing infrastructure,
and
the need for long-term management.
We based these modes on the error cases of GPG~\cite{gnu_privacy_guard}, but they are common to any software signing process based on public key cryptography.
They are: % The subsequent enumeration identifies these potential points of failure:
\begin{enumerate}
	\item \textbf{Creation Failure:}
	      The Signer does not create keys or signature files.
	      %The artifact is unsigned. %No signature is created The Signer never creates keys and signature files.
	      %Some maintainers decide not to do this, or give up on the process.
	\item \textbf{Bad Key:}
	      The Signer uses an invalid key, \eg the wrong key is used or it has become corrupted. % that does not provide the desired cryptographic guarantee, \eg based on a weak algorithm or a short key length.
	      %\JD{TODO: Our measurements focus entirely on mechanics, not on crypto strength.}
	      %The Signer may use an incorrect key or the key may become corrupted. %, or the signer may use a key that is not intended for signing.
	      %\JD{Can we distinguish (in our data) an insecure key from an incorrect key? These seem like different issues.}
	      %\TRS{Theoretically we could, but we don't have time to do that. Right now keys with a weaker alg. that are still functional still pass. I suggest cutting.}
	\item \textbf{Bad Signature:}
	      The resulting signature is incorrect or unverifiable for non-malicious reasons,
	      such as signing the wrong artifact or using an unsupported algorithm (\eg use of an unknown algorithm).
	      %This can be caused by the signer signing the wrong artifact, a software bug, or the use of an unsupported or deprecated algorithm.
	      %\JD{I don't understand why these are being grouped. ``This key/artifact pair did not make this signature'' is not quite the same as ``It is signed but the algorithm is broken''. Or maybe they are the same because a broken algorithm means the attacker can replace the package without changing the signature. Is that right? (But what is ``unsupported algorithm''?)}
	      %\TRS{Some algs. are not supported by mainline releases of GPG --- there are some closed-source ones. This is an error we can observe sometimes.}
	\item \textbf{Publishing Failure:}
	      The Signer does not publish the cryptographic material --- signature and public keys --- to locations accessible to the end user.
	\item \textbf{Key Revoked:}
	      The Signer revokes the key used to sign the artifact, \eg due to theft or a key rotation policy.
	      %Once a key is created, it can be revoked for several reasons including loss, theft, policy changes or personnel rotation.
	      %Associated signatures cannot be trusted. % be created with revoked keys, but they should not be trusted.
	\item \textbf{Key Expired:}
	      Some kinds of keys expire after a fixed lifespan.
	      Associated signatures are no longer valid. %Expired keys --- and associated signatures are no longer valid.
	\item \textbf{Signature Expired:}
	      Some signatures also expire.
	      %Once the signatures expire, they are no longer valid.
	\item \textbf{Discovery Failure:}
	      The user may fail to retrieve signatures or keys.
	      This case is distinct from Publishing Failure: the material may be available, but the user does not know where to look.
\end{enumerate}

\noindent
We omit from this list any failure modes associated with cryptographic strength (\eg short keys or broken ciphers), since these concerns vary by context~\cite{gopalakrishna2022if}.

\subsubsection{Signing Targets}
%These provide different security guarantees depending on the actors involved and the artifacts being signed.
We detail the three kinds of signing used across the four registries considered in this work.
These registries largely follow the maintainer-signer style. %, so users receive a maintainer-signed artifact.
The registries support the signing of different artifacts. %use different signing techniques.
%In this subsection, we enumerate the guarantees provided by the different signing techniques observed in our study.

\iffalse
	\JD{TODO}
	% FORMER SIGNING TARGET SECTION IS NOW CONDENSED HERE (BEGINS)
	Signing is primarily targeted at various outputs of a software development process~\cite{cooper_security_2018}.
	Signing use cases generally fall under either general Artifact signing (i.e., signing its contents) or Metadata Signing.
	% STA: removed the sentence below because we know artifacts can be anything
	%Artifact Signing can either be Firmware signing, driver signing, or Application software Signing ~\cite{cooper_security_2018}.
	Both categories generally follow the process described above with few variations.
	% In most cases as from recent research it is often advised to combine both forms of signing (i.e., signing both the software artifact and its metadata)~\cite{cappos_justin_scott_john_look_in, git-attacks}, as is the case of major Linux distributions (e.g., Debian) automotive systems (e.g., IEEE/ISTO 6100.1.0.0 ~\cite{uptane-escar}), android application repositories (e.g., F-droid~\cite{f-droid-signing}), among others.
	In most cases, recent research often advises the combination of both forms of signing, which involves signing both the software artifact and its metadata~\cite{cappos_justin_scott_john_look_in, git-attacks}. This approach is observed in major Linux distributions, such as Debian, automotive systems like IEEE/ISTO 6100.1.0.0~\cite{uptane-escar}, and android application repositories including F-droid~\cite{f-droid-signing}, among others.
\fi

\myparagraph{Maven, PyPI---Packages}
In Maven (Java) and PyPI (Python), the signing target is the software package.
%Signatures on these registries use the Pretty Good Privacy (PGP) tool.

\iffalse
	%PGP closely follows the process depicted in~\cref{fig:good_signing}. %has a fairly simple operating procedure:
	%1) user A creates a key pair,
	%2) signs a message with their private key,
	%3) sends the message to user B,
	%4) user B verifies the message using user A's public key.
	Depending on who creates and signs the message, PGP can provide different guarantees.
	For example, on \npm the registry creates signatures for each package and end users of those packages can verify that the package was signed by the registry.
	This \textit{server-side} signing guarantees that the package was not modified after being published by the registry, but does not guarantee that the package was published by the maintainer.
	On the other hand, \maven requires that maintainers sign their packages before publishing them.
	This \textit{maintainer-side} signing guarantees that the package was published by the maintainer, and allows the user to verify that the package was not modified after being signed by the maintainer.

	This system has one key challenge: the maintainer must create and manage their own key pair.
	Signatures are useless to users if they cannot find or verify the authenticity of the public key provided by the maintainer.
	If the delivery or authentication methods are compromised, then the signature is useless.
\fi

\myparagraph{Hugging Face---Commits}
In Hugging Face (machine learning models), the signing target is the git commits that underlie the package.
Signed commits may be interleaved with unsigned ones, reducing the security guarantee of a package that combines both kinds of commits.
Hugging Face's commit-based approach means that signatures only ensure that the \textit{changes} to package artifacts are authentic.

\iffalse
	Git commit signing actually uses PGP to sign commits under the hood.
	On GitHub, users can create a key pair and upload their public key to their profile.
	When a user commits code to a repository, they can sign the commit with their private key.
	In the web interface, GitHub will display a status indicator next to signed commits to indicate the status of any signature.
	Users can also verify a signature locally by downloading the public key from the user's profile and using it to verify the signature.
	The drawback of distributing keys in this way is that users must trust GitHub to provide the correct public key for the user (\ie account compromise provides a single point of failure for creating malicious signatures).
	The other drawback of signing commits, rather than files, is that these signatures only ensure that the \textit{changes} to package artifacts are authentic.
	A mixture of signed and unsigned commits could still contain changes from unauthorized sources.
\fi

\iffalse
	\JD{This part goes in methods, I think}
	\hf has a similar system, but does not make the public keys of their users accessible.
	This means that users cannot verify the authenticity of a signature unless they can acquire the public key from another source.
	In this system, users are only shown if \hf believes that the signature is valid.
	Users have no way to verify the signature themselves or see if any suspicious behavior occurs with a user's public key.
\fi

\myparagraph{DockerHub---Packages (container images)}
In DockerHub (Docker container images), the signing target is the package, \ie the container image. %a metadata tag attached to a container.
Maintainers sign the packages, but unlike in Maven, PyPI, and Hugging Face, the cryptographic materials are stored and managed by a registry service called Notary that is run in conjunction with DockerHub.
%Docker Content Trust (DCT) is a first-class system that allows users to sign and verify the contents of a Docker image.
%DCT uses a \textit{maintainer-side} signing model, where the maintainer signs the image before publishing it.
%Delegations ensure that only authorized contributors can sign tags for a repository.
%The public keys and delegations for images are stored and managed by Notary, a service that is run in conjunction with a Docker registry.
This system provides a compromise between maintainer-signer and registry-signer:
the maintainer attests to publication of the image, but the user must trust that the Notary service is not compromised (loss of cryptographic materials).
%The end user, however, must still trust that the Notary service is not compromised.

\iffalse
	\JD{TODO}
	\ST{Nit: in this case we are signing an image manifest, we may want to clarify}
	\JD{What exactly is the maintainer signing here? Checksum of the image content?}
\fi

\section{Related Works} \label{sec:related_works}

We discuss work on software signing challenges (\cref{sec:ProblemsWithSigning}) and prior measurements of signing practices (\cref{sec:EmpiricalSigningData}).

% -------------------------------------------------------------------------------------------------------------------
% -------------------------------------------------------------------------------------------------------------------
% -------------------------------------------------------------------------------------------------------------------

\subsection{Challenges of Software Signing} \label{sec:ProblemsWithSigning}

\subsubsection{Signing by Novices} \label{sec:SigningByNovices}

%Despite being widely recommended, the use and effectiveness of signing have faced criticism and raised concerns.
Like other cryptographic activities~\cite{meng2018secure,chen2019reliable}, signing artifacts is difficult for people without cryptographic expertise.
The ongoing line of ``Why Johnny Can't Encrypt'' works, begun in 1999~\cite{whitten_why_1999}, enumerates confusion in
the user interface~\cite{sheng_why_2016,Gillian_OpenITP}
and
the user's understanding of the underlying public key model~\cite{braz_security_2006,ruoti_confused_2013},
and other usability issues~\cite{reuter_secure_2020}.
Automation is not a silver bullet ---
works by Fahl \etal~\cite{fahl_helping_2012} and Ruoti \etal~\cite{ruoti_private_2015} both
found no significant difference in usability when comparing manual and automated encryption tools,
though Atwater \etal~\cite{atwater_leading_2015} did observe a user preference for automated solutions.

\subsubsection{Signing by Experts} \label{sec:SigningByExperts}
Even when experts adopt signing, they have reported many challenges.
Some concerns are specific to particular signing tools, \eg Pretty Good Privacy (PGP) has been criticized for issues such as over-emphasis on backward compatibility and metadata leaks~\cite{latacora_pgp_2019, green_whats_2014}.
Other concerns relate to the broader problems of signing over time, \eg
key management~\cite{mouli-cloud,cooper_security_2018,newman_sigstore_2022},
key discovery~\cite{melara2015coniks, malvai2023parakeet},
cipher agility~\cite{heftrig2022poster},
and
signature distribution~\cite{signature-distribution-update-hotsec}.
Software signing processes and supporting automation remain an active topic of research~\cite{newman_sigstore_2022,merrill_speranza_2023,OpenBSD}

\subsubsection{Our Contribution}
Our work measures the adoption of signing (quantity and quality) across multiple package registries.
Our data indicates the common failure modes of software signing for the processes employed by the four studied registries.
Our data somewhat rebuke this literature, showing the importance of factors beyond perceived usability and individual preference.

\subsection{Empirical Data on Software Signing Practices}
\label{sec:EmpiricalSigningData}

Large-scale empirical measurements of software signing practices in software package registries are rare.
In 2016, Kuppusamy \etal~\cite{kuppusamy_diplomat} reported that only $\sim$4\% of PyPI projects listed a signature.
%In the absence of signatures, Vu \etal found discrepancies between source code and packaged code in 65\% of the packages in PyPI~\cite{vu_lastpymile_2021}.
In 2023, Zahan \etal examined signature propagation~\cite{zahan_openssf_2023}, reporting that only 0.1\% (NPM) and 0.5\% (PyPI) of packages publish the signed releases of their packages to the associated GitHub repositories.
In 2023, Jiang \etal found a comparably low signing rate in Hugging Face~\cite{Jiang2023PTMReuse}.
In 2023, Woodruff reported that signing rates in PyPI were low, and that many signatures were of low quality (\eg unverifiable due to missing public keys)~\cite{woodruff_pgp_2023}.
Our study complements existing research by aggregating and comparing the prevalence of signing across various software package registries, in contrast to previous studies that primarily focus on single registries.
We publish the first measurements of signing in Maven and DockerHub, and the first longitudinal measurements in Maven, DockerHub, and Hugging Face.
Our multi-registry approach allows us to both observe \textit{and infer the causes of} variation in signature quantity and quality.

\section{Signing Adoption Theory} \label{sec:signing_adoption_theory}

%The open-source community has continuously noted \JD{The people noting this are researchers, not the OSS community. Maybe another verb.} a poor uptake in software signing~\cite{relations_pgp, stufft_why_nodate, woodruff_pgp_2023}.

Although software signing is recommended by engineering leaders (\cref{sec:Signing}), prior work shows that signing remains difficult (\cref{sec:ProblemsWithSigning} and successful adoption is rare (\cref{sec:EmpiricalSigningData}).
To promote the successful adoption of signing, we must understand what factors influence maintainers in their signing decisions. %do or do not sign their packages.
Even though using security techniques like signing is generally considered good practice~\cite{security_technical_advisory_group_software_2021,slsa}, maintainers do not always follow best practices~\cite{brown_digital_nudges_2021}.
Prior work has focused on the \textit{usability} of signing techniques (\cref{sec:related_works}), but we posit that a maintainer's \textit{incentives} to sign are also important.

Behavioral economics examines how incentives influence human behavior~\cite{deci_effects_1971}.
Economists typically distinguish between incentives that are intrinsic (internal) and extrinsic (external)~\cite{baddeley_behavioural_2017}.
Incentives can change how individuals make decisions, although the relationships are not always obvious~\cite{kamenica_behavioral_economics_2012,gneezy_pay_2000}.
For example, Titmuss~\cite{titmuss_gift_2018} found that paying blood donors could reduce the number of donations due to a perceived loss of altruism.
In a similar way, we theorize that incentives influence how maintainers adopt software signing.

\begin{figure}
	\centering
	\includegraphics[width=.9\linewidth]{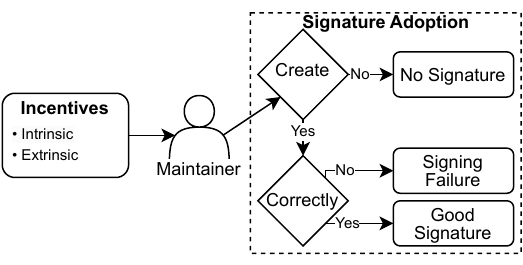}
	\caption{
		Incentives influence how maintainers adopt software signing.
		The maintainer decides weather or not to create a signature.
		If a maintainer decides to create a signature, they can either follow the signing process (\ie \cref{fig:good_signing}) correctly or not.
		Correctly following the signing process results in a good signature but incorrectly following the process results in a signing failure.
	}
	\label{fig:signing_adoption}
\end{figure}

In \cref{fig:signing_adoption}, we illustrate how incentives might apply to signature adoption.
We define a \textit{signing incentive} as a factor that influences signature adoption.
Although intrinsic incentives might contribute to signature adoption (\eg altruism), we focus on extrinsic incentives because they are more easily observed.
As summarized in~\cref{tab:signing_incentives}, we examined four kinds of external incentives  that might influence a maintainer's signing practices.
We formulated hypotheses as to their effects, but behavioral economics suggests that even the most obvious hypothesis must be tested.
%These include high profile cyber attacks, platform changes, or the publication of new security standards by trusted organizations (\eg through NIST, CISA, NSA, ENSIA, etc.).

{
% Table to show different factors that influence incentives to sign
\renewcommand{\arraystretch}{1.3} % arrows typeset ugly, stretch the rows
\begin{table}
	\centering
	\caption{
		Kinds of incentives considered.
		The second column is our hypotheses: whether the incentive was predicted to increase ($\uparrow$) or decrease ($\downarrow$) signature adoption (cf.~\cref{sec:research_questions}).
		The third column is the observed effect (cf.~\cref{sec:results}).
	}
	\label{tab:signing_incentives}
	\begin{tabular}{ccc}
		\toprule
		\textbf{Factor}   & \textbf{Expectation} & \textbf{Observed Effect} \\
		\toprule
		Registry policies & $\updownarrow$       & $\updownarrow$ quantity  \\
		%Signing Restrictions  & $\downarrow$                 \\
		Dedicated tooling & $\uparrow$           & $\uparrow$ quality       \\
		Signing events    & $\uparrow$           & None                     \\
		%Organizational Policy & $\updownarrow$ & Not measured              \\
		High startup cost & $\downarrow$         & $\downarrow$ quantity    \\
		\bottomrule
	\end{tabular}
\end{table}
}

To operationalize the concept of signing practices,
we define \textit{signature adoption} as a maintainer's decision to
(1) create a signature (quantity), and
(2) to follow the signing process correctly in doing so (quality).
To secure software supply chains, we want to identify incentives that sway the behavior of the maintainer community, not just individuals. % --- the quantity and quality of signatures.
Going forward, we thus call
\textit{Signing Quantity} the number or proportion of signed artifacts present within a given registry,
and
\textit{Signing Quality} the condition of the signatures which are present (\ie how many of them are sound, and how many display which of the failure modes indicated in~\cref{fig:good_signing}).
For example, suppose that 90\% of a registry's artifacts are signed, but only 10\% of these signatures have available public keys.
We would consider this registry as experiencing high quantity, but low quality, signing adoption.
% SB (10/15/23): I like this framing of quality and quantity of signatures. 

\iffalse
	As shown in \cref{fig:signing_adoption}, before maintainers create signatures for their software artifacts, they must first decide to do so.
	This decision is influenced by the incentives they face.
	Their organizations, if any, do not always have structured methods for deciding which security policies and products to adopt~\cite{european_union_agency_for_cybersecurity_good_2023}.
	Furthermore, the platforms they use either require signatures (such as Maven Central) or leave them as optional security features (such as Hugging Face).
	In the latter case, maintainers must decide if signatures are worthwhile to implement.
	\JD{I don't really understand the next sentence}
	Unfortunately, it is difficult to perform a cost-benefit analyses on practices like signing because the impacts are not readily measured~\cite{butler_saem}.

	Even when maintainers decide to sign their software artifacts, they may not follow the signing process correctly.
	Once again, incentives influence how maintainers follow the signing process.
	For example, maintainers may not follow the signing process correctly because they are not required to, or because there are no enforcement mechanisms in place to ensure compliance.
	In other cases, they may not follow these processes correctly because they do not consider the consequences of doing so.
\fi

Given evidence to support the theorized relationship between these factors and signature adoption (\cref{fig:signing_adoption}), one could predict the quantity and the quality of signatures in a given environment, as well as the effect of an intervention affecting maintainers' incentives.
\section{Research Questions} \label{sec:research_questions}

% Package registries are a key component in software supply chains.
% They provide downstream users with accessible software packages that reduce the engineering burden of software development.
% Unfortunately, malicious actors have increased efforts to compromise upstream components with the intent to exploit downstream users.
% Because of this, maintainers have an increased responsibility to secure their software supply chains.

% Software signing is one technique that can be used to increase trust between package registries and maintainers --- increasing security in the supply chain.
% For this technique to be practically useful, platforms and users alike must choose to adopt it.
% This means that a critical number \JD{I don't understand this language about critical number. Is there a ``critical mass'' effect like in a social network or a nuclear reactor? It's hard to understand what is intended. I suggest we cut this concept.} of packages must be signed to a high level of quality.
% In other words, most of the packages on a given registry should be signed with verifiable signatures.
% This is not the current state of signing.

\noindent
We ask three questions across two themes:

\vspace{0.1cm}
\noindent
\textbf{Theme 1: Measuring Signing in Four Package Registries}
%\vspace{0.1cm}

\noindent
Theme 1 will update the cybersecurity community's understanding on the adoption of software signing.

\begin{itemize}
	\item \textbf{RQ1:} What is the current quantity and quality of signing in public registries?
	\item \textbf{RQ2:} How do signing practices change over time?
	      %\item \textbf{RQ3:} What is the current quantity and quality of signing in public registries?
\end{itemize}

\vspace{0.1cm}
\noindent
\textbf{Theme 2: Signing Incentives}
\vspace{0.1cm}

\noindent
Theme 2 evaluates hypotheses about the effects that several types of external incentives have on software signing adoption.

\begin{itemize}
	\item \textbf{RQ3:} How do signing incentives influence signature adoption?
\end{itemize}

\noindent
To evaluate RQ3, we test four hypotheses.

%Prior measurements have provided signing quantities for PyPI~\cite{kuppusamy_diplomat} and Hugging Face~\cite{Jiang2023PTMReuse}, but rarely report quality~\cite{woodruff_pgp_2023}.

\iffalse
	\JD{The following sentence should be a backreference, probably to subsection 3.2, which should then present data for each of the registries we consider (plus others as available, \eg Nuget, rubygems, gopm, etc etc. Should we have an enumeration of registries of interest in \$3?.)}
\fi

\begin{tcolorbox} [width=\linewidth, colback=yellow!30!white, top=1pt, bottom=1pt, left=2pt, right=2pt]
	\textbf{H$_1$}:
	\textit{Registry policies that explicitly encourage or discourage software signing will have the corresponding direct effect on signing quantity}.
	This hypothesis is based on our experience as software engineers, ``reading the manual'' for the ecosystems in which we operate.
\end{tcolorbox}

\begin{tcolorbox} [width=\linewidth, colback=yellow!30!white, top=1pt, bottom=1pt, left=2pt, right=2pt]
	\textbf{H$_2$}:
	\textit{Dedicated tooling to simplify signing will increase signing adoption (quantity and quality).}
	The basis for this hypothesis is the prior work showing that signing is difficult and that automation can be helpful~\cref{sec:related_works}.
\end{tcolorbox}

\begin{tcolorbox} [width=\linewidth, colback=yellow!30!white, top=1pt, bottom=1pt, left=2pt, right=2pt]
	\textbf{H$_3$}:
	\textit{Cybersecurity events such as cyberattacks or the publication of relevant government or industry standards will increase signing adoption (quantity and quality).}
	The basis for this hypothesis is the hope that software engineers learn from failures and uphold best practices, per the ACM/IEEE code of ethics~\cite{gotterbarn1997software}.
\end{tcolorbox}

\begin{tcolorbox} [width=\linewidth, colback=yellow!30!white, top=1pt, bottom=1pt, left=2pt, right=2pt]
	\textbf{H$_4$}:
	\textit{The first signature is the hardest, \ie after a package is configured for signing adoption, it will continue to be signed.}
	Like H$_2$, this hypothesis is based on prior work showing that signing is difficult --- but knowing that the cost of learning to sign need be paid only once.
\end{tcolorbox}

% % OLD RQs from the course project
% We hope to assess the prevalence and quality of signing in open-source repositories for traditional and machine learning software. 
% As a result, we ask the following research questions:

% \begin{itemize}
%     \item \textbf{RQ1:} What is the quality of the signatures for packages on Maven Central?
%     \item \textbf{RQ2:} How many commits employ GPG signing on Hugging Face?
% \end{itemize}

% -------------------------------------------------------------------------------------------------------------------
% -------------------------------------------------------------------------------------------------------------------
% -------------------------------------------------------------------------------------------------------------------
\section{Methodology} \label{sec:methods}

\iffalse
	RQ1 and RQ2 are answered by direct measurement, describing signing practices in several registries.
	To answer RQ3, we use a quasi experiment.
	Since RQ3 affected which registries were considered.
\fi

\begin{figure}[ht]
	\centering
	\includegraphics[width=.7\linewidth]{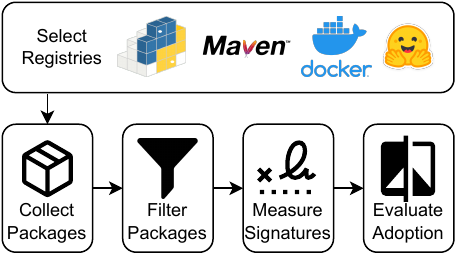}
	\caption{
		First, we select package registries that represent a range of software types and signing policies.
		Our selected registries include \registries.
		Next, we collect a list of packages for each platform.
		Then, we filter the list of packages to a sample of packages for each platform.
		On the remaining packages, we measure the quality and quantity of signatures.
		Finally, we use these measurements to evaluate factors influencing adoption.
	}
	\label{fig:methods}
\end{figure}

This section describes our methods. % we use to answer the RQs presented in in \cref{sec:research_questions}.
In \cref{sec:methods_experimental_design} we give our experimental design.
Following that design, our methodology has five stages.
As illustrated in~\cref{fig:methods}:

\begin{enumerate}
	\item \textbf{Select Registries}(\cref{sec:methods_select_registries}): Picking appropriate registries for our study.
	\item \textbf{Collect Packages}(\cref{sec:methods_collect_packages}): Methods used to collect a list of packages from each registry.
	      % SB (10/15/23): The method we use to subselect packages from a repository should be mentioned here. 
	\item \textbf{Filter Packages}(\cref{sec:methods_filter_packages}): Filter list of packages so that remaining packages are adequately versioned.
	\item \textbf{Measure Signature Adoption}(\cref{sec:methods_measure_signatures}): Measuring the quantity and quality of signatures for each registry.
	\item \textbf{Evaluate Adoption Factors}(\cref{sec:methods_evaluate_adoption}): Comparing signature adoption among registries and across time.
\end{enumerate}

% provides an overview of these stages and shows how each of them interconnect.

%In subsequent subsections  we provide enhanced detail on each stage of our overall methodology. 

\subsection{Experimental Design}
\label{sec:methods_experimental_design}

%In this subsection, we discuss the overall design of the study used in this paper.
We first describe the types of experiments needed to answer our RQs (\cref{sec:methods_answering_rqs}).
Then we summarize quasi experiments and why they work well for our study (\cref{sec:methods_quasi_experiments}).

\subsubsection{Answering our RQs} \label{sec:methods_answering_rqs}

To answer RQ1, we measure the quantity and quality of signatures in the registries of interest.
To answer RQ2, we examine trends in signature adoption and failure modes over time.
%In~\cref{sec:Background} we explained the need for these measurements.
The measurements themselves are fairly straightforward, following prior work on software signing quantity~\cite{kuppusamy_diplomat, Jiang2023PTMReuse, zahan_openssf_2023} and quality~\cite{woodruff_pgp_2023}.
In~\cref{sec:methods_measure_signatures} we define quality systematically, but our approach resembled prior work.

RQ3 requires us to test the correlation between incentives and signature adoption (quantity and quality).
Our goal is similar in spirit to the usability studies performed by Whitten \& Tygar~\cite{whitten_why_1999}, Sheng \etal~\cite{sheng_why_2016}, and Routi \etal~\cite{ruoti_confused_2013} --- like them, we are interested in factors influencing the adoption of signing.
However, we are not interested in the usability of individual signing tools, in which context controlled experiments are feasible.
Instead, we want to understand the factors affecting signing practices across package registries.
%and must perform an experiment to test our hypotheses.
A controlled experiment %true experiment to test our hypothesis and answer our RQs.
would randomly assign maintainers to platforms with different registry factors, hold other variables constant, and measure the effect for hypothesis tests.
%Subsequent differences in signature adoption would provide us with evidence to support or refute our hypotheses.
As this level of control is impractical,  we instead use a \textit{quasi experiment} to answer RQ3.

\subsubsection{Quasi Experiments} \label{sec:methods_quasi_experiments}
\textit{Quasi experiments}
\footnote{
	% \JD{I think this paragraph could be compressed into a 2 sentence footnote after the first mention of ``quasi experiment''. WDYT?}
	% \TRS{PTAL I did it in 3.}
	The terms \textit{quasi experiment} and \textit{natural experiment} are often used interchangeably, but some researchers distinguish between the two.
	Treatments applied in a natural experiment are not intended to influence the outcome, whereas treatments in a quasi experiment are planned~\cite{remler_research_2015}.
	Since some registry factors are intended to cause changes in signature adoption, our study is a quasi experiment, not a natural experiment.
}
are a form of experiment whereby
(1) \textit{treatments} (instances of independent variables) are applied to \textit{subjects};
and
(2) \textit{outcomes} (dependent variables) are measured;
but
(3) the assignment of treatments to subjects is not random~\cite{felderer_contemporary_2020, remler_research_2015}.
Instead, the application of treatments is based on characteristics of the subjects themselves~\cite{wohlin_experimentation_2012}.
This method is often used where controlled experiment is infeasible, \eg measuring impacts of government policies on populations~\cite{remler_research_2015, wooldridge_introductory_2013, meyer_natural_1995}.
These studies produce the strongest results when treatments occur independently of the subjects, or are \textit{exogenous}~\cite{meyer_natural_1995, remler_research_2015}.
% \JD{Define exogeneous treatments here in this paragraph, where you are explaining terms and the things the reviewer should be checking for.}

% \JD{Should this paragraph be the second paragraph in this subsubsection? Then the order would be ``This is a Quasi Experiment ; This is why it's a good fit for us; This is how we build a Quasi Experiment''}
A quasi experiment would allow us to test our hypotheses by leveraging naturally occurring differences between registries.
If we can identify registries that vary along the dimensions of interest, then comparing signing adoption in these registries would allow us to infer cause-effect relationships from the incentives of interest.
%Such a comparison is commonly referred to as a \textit{difference-in-differences} because of the combination of before-after comparisons and comparisons between groups~\cite{remler_research_2015}.
Furthermore, the existence of multiple time periods and comparison groups also strengthens the outcomes of this approach~\cite{meyer_natural_1995}.
% SB (10/15/23): I like this description and justification for quasi experiments. 

Our quasi experiment relies on a major assumption:

\begin{tcolorbox} [width=\linewidth, colback=blue!10!white, top=1pt, bottom=1pt, left=2pt, right=2pt]
	\textbf{Assumption of Structural Similarity}:
	Our quasi-experiment assumes that there are no uncontrolled confounding factors.
	Such factors would differentially affect software signing adoption between registries.
	In other words, we assume that the registries are used by software engineers in similar ways, without registry-specific signing influences other than those considered.
\end{tcolorbox}

We base this assumption in the cybersecurity and software engineering research literature.
For example, Zahan \etal~\cite{zahan2023software} compared NPM and PyPI and found comparable behaviors across a range of security measures and practices, including security policies, vulnerabilities, dependency update tools, maintenance procedures, signed-releases, and potentially risky workflows.
Bogar \etal~\cite{bogart_2021} offered further support for this assumption in a study of several software ecosystems, writing that ``\textit{Ecosystems tend to share many values but differentiate themselves based on a few distinctive values strongly related to their purpose and audience.}''
While there is limited research on PTM registries, Jiang \etal's~\cite{Jiang2023PTMReuse} reported that the PTM re-use workflow for PTMs on \hf to that of traditional software.
This suggests comparability between the PTM space and the traditional software ecosystem space.
These and other works suggest a level of uniformity across registries with respect to engineering practices. %, supporting our assumption. %, indicative of consistent behavior among their respective actors.
%All registries support reuse.

With this assumption in mind, we conduct the quasi experiment.
In our study, registry factors are \textit{treatments}, maintainers are \textit{subjects}, and signature adoption is the \textit{outcome}.
As noted, applying random registry factors to maintainers is impractical --- this is effectively the same as randomly assigning maintainers to different platforms.
Instead, characteristics of the maintainers (the registries they naturally use) determine which registry factors they experience.
Since maintainers' registry selections are not known to be influenced by signing infrastructure or policy, we can also consider registry factors as exogenous treatments.

We illustrate our experimental design with two evaluations, which we perform later in detail.
Our four hypotheses for RQ3 include incentives that are registry-specific (H$_1$, H$_2$) and that are registry-independent (H$_3$, H$_4$).
To assess the effect of registry-specific incentives, we examine whether the date of an associated event is correlated with a significant change in signing adoption within only the pertinent registry.
For example, if a change in PyPI's signing policy changes the signing rates in PyPI, while rates in other registries are undisturbed, then we would view this as support for H$_1$.
Conversely, to assess the effect of registry-independent incentives, we examine whether the date of an associated effect is correlated with a significant change in signing adoption in multiple registries.
For example, if after a major software supply chain attack we see changes in signing adoption in PyPI and Maven, then we would view this as support for H$_3$.

\vspace{0.1cm}
\noindent
Based on this design, we now proceed through the five stages indicated in~\cref{fig:methods}.

\subsection{Stage 1: Select Registries} \label{sec:methods_select_registries}
% \JD{Look in the TEX -- I made this one sentence (keeping the one with more casual language) instead of 2 sentences (with more formal language)}
%In this section, we discuss the selection requirements (\cref{sec:methods_selection_requirements}) and the registries actually selected for use in this study (\cref{sec:methods_selected_registries}).
In this section, we explain what makes a registry a good candidate for this study (\cref{sec:methods_selection_requirements}) and justify our use of \registries (\cref{sec:methods_selected_registries}).

\subsubsection{Selection Requirements} \label{sec:methods_selection_requirements}

We searched for software package registries which have natural variations in signing incentives.
We selected some registries that have experienced changes to their signing infrastructure and policies over time.
We focused on package registries with maintainer-signed signatures, as defined in~\cref{sec:Background}, for two reasons:
1) they place a greater burden on the maintainer and thus the effect of incentives would be more observable; and
2) they provide a better guarantee of provenance than server-signed signatures (\ie the artifact hasn't been modified between the maintainer signing it and uploading it to a registry).
% \JD{Also they provide a better security guarantee, right? I think we should describe the positive part too.}
%We wanted package registries that package different kinds of software, to 
Finally, the selected registries should be popular so that they are representative of publicly available software.

%For this reason, we choose to select multiple registries with characteristically different signing infrastructure and policies (\ie each registry has a different combination of signing difficulty and incentive) ---
%this allows us to perform corss-sectional comparisons.
%this allows us to perform longitudinal comparisons.

% \JD{I think you should collapse this paragraph into one sentence that occurs at the end of the first paragraph.}
% \Kelechi{yes, this still sounds like the first paragraph, maybe merge them}
% Preferably, our registry selection should also generally represent open-source software.
% This means that we should select registries which host software artifacts from a variety of software domains.
% For example, we should not select two registries which host only Python packages.
% Instead, we should select one registry which hosts Python packages and another which hosts machine learning models.
% This ensures that our results are not specific to a single software domain.

\subsubsection{Selected Registries} \label{sec:methods_selected_registries}

% \begin{figure}
% 	\centering
% 	\includegraphics[width=.9\linewidth]{fig/registry_matrix_v1.png}
% 	\caption{Matrix showing the relative difficulty and incentive of selected software registries. Arrows show a change in difficulty or incentive due to a change in the platform.
% 		% TODO: Update pypi and docker locations
% 		\TODO{Update pypi and docker locations}
% 		% TODO: Add more 'useful' pixels.
% 		\TODO{Add more 'useful' pixels.}
%   \SB{I am thrown by "due to a change in the platform". We mean due to evolution over time?}
% 	}
% 	\label{fig:registry_matrix}
% \end{figure}

% \SB{Since Hugging Face has no signatures, it cannot serve as the only model repository for our study. How about trying others like ModelZoo, Paperswithcode?}
% \TRS{There are signatures in hugging face - the adoption rate is just so low that a small enough sample size is not representative. Our new data shows this.}
% \JD{Do we have a sense of adoption in the other model repositories? It would be helpful to tackle this criticism with data if we can.}
% \TRS{
% 	\begin{itemize}
% 		\item ModelZoo stores its models on github --- some of these use git commit signing. We could use this as a comparison group for hugging face, my breif look at these show that adoption rates are likely higher.
% 		\item paperswithcode also stores models on github --- the story is the same as model zoo.
% 	\end{itemize}
% }
% \ST{consider that github commit signing can be performed on the server side with the GH key. I suspect the bulk of signatures here will be those}
Following these requirements, we identified \numRegistries registries for study: \registries.
\cref{tab:registries} summarizes these registries. % by listing the name, software type, and signature type. %, \textit{difficulty}, and \textbf{incentive} of each registry.
They represent some of the most popular programming languages~\cite{cass2023top} (Java, Python),
the most popular container technology~\cite{pahl2017cloud} (Docker),
and
the most popular ML model hub~\cite{jiang2022PTMSupplyChain} (Hugging Face).
Next we elaborate on each selected registry.
All data is as of April 2024.

\begin{table}[ht]
	\centering
	\small
	\caption{
		The selected package registries, their associated software type, and signature type.
	}
	\begin{tabular}{lll}
		\toprule
		\textbf{Registry Name} & \textbf{Software Type} & \textbf{Signature Type} \\ \toprule
		PyPI                   & Python                 & PGP (now deprecated)    \\
		Maven Central          & Java                   & PGP                     \\
		Docker Hub             & Containers             & DCT                     \\
		Hugging Face           & ML Models              & Git Commit Signing      \\
		\bottomrule
	\end{tabular}
	\label{tab:registries}
\end{table}

%We selected these registries because they provide natural variation between signing \textit{difficulty} and \textbf{incentive} across multiple software domains.

% \Kelechi{A. IMO Your baseline definition (medium and neutral) is good but readers need an exhaustive picture of what your spectrums for difficulty and incentive look like. A figure would be helpful here. The figure could take the form of a scale or even a table, showing easy, medium and difficult. This will also save a lot of explanation in text. This could also be merged with fig 4.}

% \JD{Everything (or almost everything) in these paragraphs should be parallel in construction --- report the same information in the same order.}
% Order here shoudld be: name, what is it?, software type, signature type, difficulty, incentive

\myparagraph{(1) PyPI}
\pypi is the primary registry for the exchange of software packages written in the Python programming language.
\pypi hosts more than 520,000 packages~\cite{pypi_website}.
\pypi allows maintainers to sign packages with PGP signatures.
The \pypi registry owners deemphasized the use of PGP signatures on \pypiPGPDeemp~\cite{stufft_pgp_2018} and later deprecated them on \pypiPGPRemoved~\cite{stufft_removing_2023}.
Pre-existing signatures remain, but users cannot (easily) add new signatures.

\myparagraph{(2) Maven Central}
Maven Central (Maven) is the primary registry for the exchange of software packages written in the Java programming language.
Maven hosts more than 499,000 packages~\cite{ecosystems_website}.
Like \pypi, Maven allows maintainers to sign packages with PGP signatures.
%popular development tool for Java projects that streamlines the build process, assists with dependency management, and helps with deployment.
% \maven is the default Maven repository~\cite{maven_central}.
% \maven hosts open-source Java artifacts that can be easily integrated into Maven projects.
%
%Maven allows 
Unlike \pypi, on Maven, signatures have been mandatory since 2005~\cite{relations_pgp}.
%PGP signatures are relatively difficult to create, manage and upload (see \cref{app:signing_tools_pgp}) compared to Git commit signing, so we classify this registry as having \textbf{hard difficulty}.
%Since \maven requires signatures, maintainers must sign their packages if they want to publish them --- this is an especially strong incentive \textbf{positive incentive}.

% Since \maven uses PGP signatures, we classify this registry as having hard difficulty.
% As noted in the \pypi description, PGP signing is more difficult than signing git commits (our baseline) because maintainers must manually create signature files and manage their own keys.
% 
% We classify \maven's incentives as positive.
% Since \maven requires signatures, maintainers must sign their packages if they want to publish them.
% This policy is essentially the opposite of \pypi's policy and is a strong incentive for maintainers to sign their packages.

\myparagraph{(3) Docker Hub}
\docker is the primary registry for the exchange of virtualized container images in the Docker format.
\docker hosts more than 1,000,000 container images~\cite{ecosystems_website}.
\docker uses Docker Content Trust (DCT)~\cite{docker_dct} to sign container images.
As noted in~\cref{sec:Background}, \docker has dedicated tooling for signing, integrated in the Docker CLI. % and requires no external tooling to use.
Sigstore's Cosign also supports signing docker images~\cite{lorenc_cosign_2021}, but is not yet well integrated in \docker.
In 2019, \docker began including more information about image provenance (\eg author, OS, digest, architecture) on its web UI, but does not mandate signing like Maven does.
% Another relevant event for \docker was the addition of support from Cosign on \dockerCosignAdded~\cite{lorenc_cosign_2021}. %, although this feature is reportedly not well established (see~\cite{app:signing_tools_sigstore}).

\iffalse
	%(see \cref{app:signing_tools_dct}).
	This makes DCT marginally easier to use than Git commit signing, so we classify this registry as having a \textbf{slightly easy difficulty}.
	\docker encourages maintainers to sign tags by allowing users to filter images by signature status.
	This is not as strong of an incentive as \maven's requirement, so we classify this registry as having a \textbf{slightly positive incentive}.
\fi

\myparagraph{(4) Hugging Face}
\hf is the primary registry for the exchange of neural network models~\cite{jiang2022PTMSupplyChain}.
% \JD{At the end of this sentence you need to cite Wenxin's SCORED paper that shows this}
\hf hosts more than 590,000 models~\cite{hugging_face_website}.
% \JD{as of MONTH YEAR}
\hf supports the signing of git commits~\cite{hugging_face_security}.
\hf has no stated policy towards signing, and we are aware of no signing events specific to \hf.

\subsection{Stage 2: Collect Packages} \label{sec:methods_collect_packages}

Next, our goal was to collect a list of the packages from each of the selected registries, so that we could sample from it and measure signing practices.
For each registry, we attempted to enumerate all packages available on the platform.
We used ecosyste.ms~\cite{ecosystems_website} as an index for the packages available from \maven and \docker.
For \hf, we used the \hf API to collect packages.
For \pypi, we used the Google BigQuery dataset~\cite{pypi_bigquery} to collect packages.
% \JD{CITE THAT DATASET HERE}
In the remainder of this subsection, we describe the package structure and collection techniques used for each registry.

\subsubsection{The Ecosyste.ms Cross-registry Package Index}
Ecosyste.ms provides a comprehensive cross-registry package index, aggregating package data from multiple registries into a database.
Periodically, ecosyste.ms releases datasets that can be downloaded and subjected to detailed queries.
In this work, we used the most recent version of the dataset --- \ecoDate. %, which represents the most up-to-date release at the time of our writing.
%Ecosyste.ms had the neproved invaluable in sourcing data for the Docker and Maven registries, it did not encompass the full spectrum of information we required, particularly for Maven.
% This dataset was missing some information from Maven, and did not index the Hugging Face registry.
This dataset indexed \pypi, \docker, and \maven, but not \hf.
We did not use this dataset for \pypi because the BigQuery dataset contains signing information.

We sought to obtain data up to \finishDate from each registry.
We arbitrarily set the start date to \startDate imitating~\cite{woodruff_pgp_2023}, so that the reported data would not be too different from current practices.

%To bridge this data gap, we directly accessed the Maven registry to obtain the requisite package-level data.

\subsubsection{Details Per-Registry}
\myparagraph{PyPI}
In \pypi, packages are distributed by version as \textit{wheels} or \textit{source distributions}
(\ie each package may have multiple versions each with their own distributions).
For example, the latest version of the \textit{requests} package is \textit{2.31.0} (as of this writing) and has two distribution files:
(1) a wheel file named \textit{requests-2.31.0-py3-none-any.whl} and
(2) a source distribution file named \textit{requests-2.31.0.tar.gz}.
Both of these files can be signed, so we collect each of these files during our assessment of \pypi.

To collect all packages from \pypi between \startDate and \finishDate, we used Google's BigQuery \pypi dataset.
This dataset contains a list of package distributions and associated metadata for each package hosted on \pypi.
% To collect all packages from \pypi between \startDate and \finishDate, we use Google's BigQuery \pypi dataset.
% This dataset contains a list of package distributions and associated metadata for each package hosted on \pypi.
% One of the metadata fields is \textit{has\_signature} which indicates whether or not a package has a signature.

Since this study is only concerned with the quality and quantity of signatures, we did not need to download any packages without signatures (we only need to count how many packages have no signature).
We downloaded signed packages to assess the quality of their signatures.

% \myparagraph{NPM}
% % TODO: Add NPM
%
% To collect all packages from \npm between \startDate and \finishDate, we used ecosystem's data dump from 22 Oct 2023.
% % SB (10/15/23): How can a data dump from 8 Aug 23 give us the packages from \startDate?
% This data dump contains a list of package distributions and associated metadata for each package hosted on \npm.

\myparagraph{Maven Central}
\maven packages are stored in a directory structure
%\footnote{\url{https://repo1.maven.org/maven2/}}
organized by namespace, package name, and version number.
Each version of a package contains several files for use by the downstream user.
These typically include $.jar$, $.pom$, $.xml$, and $.json$ files which include the package, source distributions, tests, documentation, or manifest information.
Each of the files included in a package version typically has a corresponding PGP signature file with a $.asc$ extension.

We used ecosyste.ms' data dump from \ecoDate to collect packages from \maven.
This data dump contains a list of package distributions and associated metadata for each package hosted on \maven.

%\JD{CRITICAL: This is a sample of 5K packages instead, yes?}

% To collect all packages from \maven between \startDate and \finishDate, we recursively iterate through the maven repository and collect packages which were last updated in our date range.
% The directory structure of \maven also allows us to see the dates associated with each file uploaded to the registry.
% As a result, we are able to collect and sort files from \maven by date.

% \maven packages exist in a common repository located at , but a more user-friendly interface exists at \url{https://central.sonatype.com/}. 
% The former provides a simple directory structures whereas the latter provides more user-focused information.
% Since the packages can be easily iterated over, we decide to assess all packages on the platform.
% As we will note in \cref{sec:methods_signature_evaluation}, public keys are typically stored in public key servers, but some of them are not.
% For this reason, we also re-evaluate a subset of packages without keys made available on public key servers.
% \TODO{This part might need some tweaking once we find out how many there are.}
% We sample a random \mavenReevaluateSelected packages without keys on public servers to re-evaluate.

\myparagraph{Docker Hub}
\docker packages are organized into \textit{repositories} which are collections of \textit{images}.
Each repository contains \textit{tags} which are versions of the image.
These tags can be signed using Docker Content Trust (DCT).
DCT is a Docker-specific signing tool built on Notary~\cite{notary}. % under the hood.

To collect all packages from \docker between \startDate and \finishDate, we use ecosyste.ms' data dump
\JD{Taylor, please place the following in one of the READMEs in your github artifact, but not here in the paper:
	We use \url{https://ecosyste.ms/} to get this list, however, after a review of their API source, we found that their list of packages is built by continually appending a ``list'' with the most recently uploaded packages from Docker Hub.
	This may lead to small discrepancies between what is on the ``list'' and what is on \docker.}
from \ecoDate.
This data dump contains a list of package distributions and associated metadata for each package hosted on \docker.
% \docker implements the Docker HTTP Registry API V2~\cite{docker_api_v2}.
% Unfortunately, the instance of this API running on \docker does provide a catalog of all the repositories available.
% To get this list, we use a third-party API~\footnote{
% 	We use \url{https://ecosyste.ms/} to get this list, however, after a review of their API source, we found that their list of packages is built by continually appending a ``list'' with the most recently uploaded packages from Docker Hub.
% 	This may lead to small discrepancies between what is on the ``list'' and what is on \docker.
% }
% to collect a list of all repositories on \docker, then use this list in conjunction with the Docker Registry API to collect all repository tags within our collection date range.

\myparagraph{Hugging Face}
\hf hosts models, datasets, and spaces for machine learning.
For the purpose of this study, we only focus on the models, which are stored as git repositories.
\hf uses git commit signing, \ie signatures occur on a per-commit basis for each repository.

To collect all of the model packages on \hf between \startDate and \finishDate, we use the Hugging Face Hub Python interface to generate a list of packages (and metadata) in our date range.
We then iteratively clone all repositories from \hf.

\iffalse
	\bigskip

	We collect packages from each of our selected registries (\registries) between the \startDate and the \finishDate date range of our study.
	For each of the selected registries, we attempt to collect and assess all of the packages that they host (with some technical limitations).
	Each of the registries is organized differently, so the unit of analysis varies (\eg git commits vs individual files).
	We then use collected packages to evaluate signatures in \cref{sec:methods_measure_signatures}.
\fi

\subsection{Stage 3: Filter Packages} \label{sec:methods_filter_packages}

\begin{table}
	\centering
	\caption{
		Packages available after each stage of the pipeline.
		Collect packages refers to the total number of packages available in the registry.
		Filter packages refers to the number of packages with $\geq5$ versions between \startDate and \finishDate and non-gated \hf models.
		Measure signatures refers to the number of packages we attempted to measure.
		Due to download rates, we only measure a random sample of the filtered \maven packages (10\% of the total population).
	}
	\begin{tabular}{lcccc}
		\toprule
		\textbf{Stage}     & \textbf{PyPI} & \textbf{Maven} & \textbf{Docker} & \textbf{HF} \\
		\toprule
		Collect Packages   & \pypiPackages & \mavenPackages & \dockerPackages  & \hfPackages \\
		Filter Packages    & \pypiFilter   & \mavenFilter   & \dockerFilter    & \hfFilter   \\
		Measure Signatures & \pypiMeasure  & \mavenMeasure  & \dockerMeasure   & \hfMeasure  \\
		\bottomrule
	\end{tabular}
	\label{tab:filter}
\end{table}

After obtaining the lists of packages, we filtered them for packages of interest to our study.
Since our study was interested in effects over time, our primary filter was for packages with multiple versions.
In all of the registries, we filtered for all packages with $\geq5$ versions between \startDate and \finishDate.
\footnote{
	Note that for \docker we filtered for packages with $\geq5$ tags and for \hf we filtered for packages with $\geq5$ commits.
	The notion of a version for ML models is less clear than for software packages, so we used commits as a proxy.
	\hf has some mechanisms (\eg tags) for versioning, but we found that these are not consistently used.
}
\JD{Just confirming that this footnote is still accurate? Also, I recall we explored some other definitions for \hf. Please add a sentence clarifying that we did measure the use of tags, and indicate the proportion of PTMs that use them.}
\TRS{I don't have that data on hand, but I can get it.}
\JD{Too late. Please put it into the github repo instead.}

On \hf, we also filter models that are gated (\ie they have some sort of access control).
Examples include \textit{pyannote/segmentation} which requires users to agree to terms of use.
These models account for 0.9\% (\hfGated) of the total models.
We are unable to tell how many of these models have $\geq5$ commits in the time period, without gaining access.
In many cases, this requires agreeing to terms of use, which we are unable to do at scale.

See~\cref{tab:filter} for the number of packages available after each stage of the pipeline.

\subsection{Stage 4: Measure Signatures} \label{sec:methods_measure_signatures}

After filtering, we measured the quantity and quality of the signatures associated with the surviving packages.

%In this subsection, we describe our methods for evaluating the quality and quantity of signatures in each of the selected registries.
In \cref{sec:signing_adoption_theory} we defined signature quantity as  the fraction of signed artifacts in a registry, %number as the fraction of signatures present in a registry,
% SB (10/15/23): "fraction of signatures"?
and quality as the fraction of good signatures among those. %condition of the signatures present.
Since signatures apply to different units of analysis in each registry, we defined quantity and quality somewhat differently for each registry.
The nature of the signable artifacts was discussed earlier in this work.
However, certain aspects of quality cannot be measured in all registries (\eg signature expiration is not applicable to git commits).
We use the remainder of this subsection to clarify signature quality for each registry.

For quality, not all failure modes can be measured in each registry.
Recall that \cref{fig:good_signing} indicated failure modes in a typical signing scheme.
%Since some of the registries use common signing schemes, such as GPG, we are able to measure quality in a similar manner across registries.
%For example, we can measure the presence of a signature, the validity of a signature, and the expiration of a signature in a similar manner across registries.
However, some registries use unique signing schemes, such as Docker Content Trust (DCT).
These schemes do not allow us to measure quality in the same manner as other registries.
Although they use similar cryptographic methods, we cannot measure points of failure in the same manner as other registries.
Using the numbering system of~\cref{fig:good_signing}, in~\cref{tab:measurable_signature_failures} we indicate which signature failures can be measured on each platform.
%We use this table to guide our evaluation of signature quality in each registry.

% Table to show which signature failures we can measure
\begin{table}
	\centering
	\caption{
		Signature statuses and whether or not they are measurable on each platform.
		\cmark: Measurable.
		\xmark: Not measurable.
		--: Theoretically measurable.
		PK: Public key.
	}
	\begin{tabular}{lccccccc}
		\toprule
		\textbf{Status}   & \textbf{Failure \#} & \textbf{PyPI} & \textbf{Maven} & \textbf{Docker} & \textbf{HF} \\
		\toprule

		Good Signature    & --                  & \cmark        & \cmark         & \cmark          & \cmark      \\ \midrule
		No Signature      & 1                   & \cmark        & \cmark         & \cmark          & \cmark      \\
		Bad Signature     & 3                   & \cmark        & \cmark         & \xmark          & \cmark      \\
		Expired Signature & 7                   & \cmark        & \cmark         & \xmark          & --          \\
		Expired PK        & 6                   & \cmark        & \cmark         & \xmark          & --          \\
		Missing PK        & 4,8                 & \cmark        & \cmark         & \xmark          & --          \\
		Revoked PK        & 5                   & \cmark        & \cmark         & \xmark          & --          \\
		Bad PK            & 2                   & \cmark        & \cmark         & \xmark          & --          \\
		\bottomrule
	\end{tabular}
	\label{tab:measurable_signature_failures}
\end{table}

\subsubsection{PGP} \label{sec:pgp}

\pypi and \maven use PGP signatures to secure packages.
The measurements for these registries are similar, so we describe commonalities here.

\myparagraph{Discovery}
For PGP, we need to discover the public keys associated with each signature.
The most common method for sharing public PGP keys is to use a public key server.
Sonatype recommends the Ubuntu, OpenPGP, and MIT servers~\cite{sonatype_pgp}.
We found 5 more servers via Google searches.

On this set of 8 servers, we conducted a small-scale experiment to determine which servers to use in our study.
Using a sample of $\sim$3,800 keys from \pypi and \maven, we found that only 4 servers worked reliably (\ie some servers are no longer functional or perform slowly).
For example, the famous \textit{pgp.mit.edu} server often times out.
% \JD{Add `(a sample of DETAILS)', but I think we should also move this to the end of the paragraph, eg saying `we started with 7 but in a small experiment we found that 4 of them contained 95\% of all keys we could find' }
% \JD{I don't quite understand what `well known/easily searchable' means here. Like, they show up on Google?}
% \JD{what does `reliabliity' mean?}

% \JD{The following is a bit confusing. We refer to Sonatype that lists 3 servers above, and now we list 4 servers (including only 2 of the ones from Sonatype). Be clearer about where these came from.}
We found that four servers responded consistently:
\textit{keyserver.ubuntu.com}
\textit{keys.openpgp.org}
\textit{keyring.debian.org},
and
\textit{pgp.surf.nl}
For each key, we queried each of these servers in order to find the public key associated with the signature.
If we were unable to find the key in any of these servers, we marked the signature as having a missing public key.

Among our selected servers, the Ubuntu server was able to discover the most keys, followed by the Surf, OpenPGP, and Debian servers.
The Surf server is queried last because has the most overlap with the Ubuntu server (\ie we are more likely to find the key by looking at other servers first).

\myparagraph{Verification}
To verify a PGP signature, we use the \textit{gpg} command line utility.
This utility provides a \textit{verify} command which can be used to verify the validity of a signature.
This command returns the status of the signature verification.
We parse this status to determine the quality of a signature.

\myparagraph{Expiration}
For our measurements, we consider a key to be expired if the key's expiration had passed at the time of our measurement.
There is no definitive way to determine if a signature was created before or after the key expired.

\myparagraph{Cryptographic Algorithm}
The strength of a PGP key is determined by the cryptographic algorithm.
The \textit{gp} command line utility provides a \textit{list-packets} command which can be used to extract metadata from a signature.
This command returns the cryptographic algorithm and, particularly important for RSA keys, the key length.
We compare this to NIST's recommendations~\cite{NIST_RSA_SIZES}.

\subsubsection{Details Per-Registry}
\myparagraph{PyPI}
\pypi uses PGP signatures to secure packages.
As a result, we use the methods described in \cref{sec:pgp} to measure the quality and quantity of signatures on \pypi packages.
% This command returns a status code which indicates whether or not the signature is valid.
% We use this status code to determine whether or not a signature is valid.

% \myparagraph{NPM}
% % TODO: Add NPM
% NPM signs packages on behalf of the user at the time of publication using a server-side held private key. This process is done transparently to the user and the signatures are kept in-band as part of the package's repository metadata (i.e., inside of its ``packument''). During the period of the study, NPM used both a GPG, and then moved to ECDSA (RFC6979).
% %\JD{CRITICAL: This is empty}

\myparagraph{Maven Central}
\maven requires PGP signatures on all artifacts.
As a result, we use the methods described in \cref{sec:pgp} to measure the quality and quantity of signatures on \pypi packages.

For \maven, we were unable to sample the entire filter population.
Due to the high adoption quantity and the number of files associated with each \maven package, verifying \maven packages requires a large amount of network traffic to download every file and its signature.
For this reason, we choose to only check a random sample of the filter population (10\% of the total population).
See~\cref{tab:filter} for the number of packages we attempted to measure versus the total number of packages available.

\myparagraph{Docker Hub}
\docker uses Docker Content Trust (DCT) to sign image tags.
% DCT is the default signing scheme for \docker.
% However, SigStore has recently updated Cosign to support \docker.
% Both of these signing tools have command line utilities which can be used to verify signatures.
% We use these utilities to verify the validity of signatures.
This tool has first-class support in the Docker CLI, so we use the \textit{docker trust inspect} command to verify the validity of signatures.
Since DCT automates much of the signature process, we can only measure the presence of a signature (\ie you cannot upload a bad signature with DCT).
For this reason, all signatures are considered valid unless they are missing.
% For DCT, we use the \textit{docker trust inspect} command to verify signatures.
% For Cosign, we use the \textit{cosign verify} command to verify signatures.

\myparagraph{Hugging Face}
\hf uses git commit signing for its models.
Typically, git commit signatures are verified using the \textit{git verify-commit} command.
This uses PGP under the hood, and still requires a public key to verify the signature.
On platforms like GitHub, the public key is stored on the user's profile.
However, \hf does not currently expose the public keys for its users~\cite{hf_forum_post}.
% \JD{In the footnote: How do we know t hey are aware of this? Did we ask them? If so, say `(personal communication)'. If it's public knowledge, add a citation.}
Since users upload keys to \hf they do not necessarily upload them to a public key server as in~\cref{sec:pgp} --- we tried and had a low discovery rate.
%
\iffalse
	\footnote{
		We could not find an appreciable number of keys for \hf using public key servers (as in \cref{sec:pgp}).
		%We experienced a large mismatch between our local verification and what was present on the \hf website.
	}
\fi
% \JD{In the preceding footnote, I suggest you add a cref back to 6.5.1 --- I think that's where the method you are referring to is described?}
For this reason, verifying signatures with \textit{git verify-commit} is not actually representative of the signature quality.
Instead, we use Hugging Face's UI to verify the validity of signatures.
Verified signatures are marked as good, and unverified signatures are marked as bad.
% To measure the adoption of selected packages on \hf, we adapted a script used by Jiang \etal to create PTMTorrent~\cite{jiang_ptmtorrent}.
% This script uses a Python wrapper for Git to download repositories locally.
% Since repositories are often large and we only care about commit signing, we download bare repositories to speed up the process.
% We then iterate through all commits in the downloaded repository and use the built-in \textit{git verify-commit} command to check for signatures on each commit.
% We log output of this command for later analysis.

\bigskip

\subsection{Stage 5: Evaluate Adoption} \label{sec:methods_evaluate_adoption}

For RQ1 and RQ2, our method is to report and analyze the statistics for each registry.

To answer RQ3, we need to evaluate hypotheses H$_1$--H$_4$.
For these hypotheses, we need distinct factors corresponding to each class of incentive.
We identified a set of factors through web searches such as ``software supply chain attack'' or ``government software signing standard''.
Four authors collaborated in this process to reduce individual bias.

The resulting set of factors are given in~\cref{tab:registry_changes}.
We identified changes in registry policies and in three kinds of signing events: software supply chain attacks, government actions, and industry standards.

Using these factors, we perform exploratory data analysis (visual analysis) to identify trends in signing adoption.
This analysis was performed on another sample of the data to avoid biasing our final statistical tests.
On the final data, we use statistical tests to evaluate hypotheses H$_1$--H$_3$.

We conduct statistical tests as follows:
\begin{itemize}
	\item We assume that if an event (incentive) impacts a given registry, then the effect will be measurable within 6 months. This time horizon was selected to permit some amount of lag, while not introducing too many possible confounding events within the time frame.
	\item The data used in the tests are the daily signing rates from each registry (calculation: Number of signed units / Total number of units).
	\item We select a 99\% confidence level (p-values should be $<$ 0.01 to be considered significant)
\end{itemize}

We note two caveats:
\textit{First}, there are many comparisons that could be performed (4 registries x 16 events in Table 5), but every additional test increases the risk of Type-1 errors (false positives).
For this reason, we only test the hypotheses that are supported by exploratory data analysis and apply a Bonferroni correction to our p-values.
\textit{Second}, where present, statistically significant results indicate correlation, not causation.
Our research is motivated by an underlying predictive theory.
If the predicted results occur at a statistically significant level, they support the theory.

We conducted 3 kinds of statistical tests: (1) a one-way ANOVA test on overall differences between registry adoption quantity; (2) a subsequent Tukey tests; and (3) selected independent t-tests to compare signing before and after selected events.
All tested distributions meet the assumptions inherent in the corresponding tests.

% \JD{Provide more information about the stats tests here.}
% \JD{Clarify that preliminary means `on a previous snapshot' so that it doesn't corrupt our final stats tests}

\begin{table}
	\centering
	%\small
	\caption{
		Factors that could provide signing incentives.
		Factors were identified via web searches by four authors, and categorized by incentive type per~\cref{tab:signing_incentives}.
	}
	\begin{tabular}{llc}
		\toprule
		\textbf{Factor}     & \textbf{Category}          & \textbf{Date} \\
		\toprule
		PyPI De-emph        & Registry policy            & Mar 2018      \\
		Docker Hub Update   & Registry policy            & Sep 2019      \\
		PyPI Removed PGP    & Registry policy            & May 2023      \\
		% SKS Shutdown        & Dedicated tooling          & Jun 2019      \\
		\midrule
		NotPetya            & Signing event (attack)     & Jun 2017      \\
		CCleaner            & Signing event (attack)     & Sep 2017      \\
		Magecart            & Signing event (attack)     & Apr 2018      \\
		DockerHub Hack      & Signing event (attack)     & Apr 2019      \\
		SolarWinds          & Signing event (attack)     & Dec 2020      \\
		Log4j               & Signing event (attack)     & Dec 2021      \\
		%  \midrule
		%		Codecov             & Build Tool        & Apr 2021      \\
		\midrule
		NIST code signing   & Signing event (government) & Jan 2018      \\
		CISA Publication    & Signing event (government) & Apr 2021      \\
		Exec. Ord. 14028    & Signing event (government) & May 2021      \\
		\midrule
		CMMC                & Signing event (standard)   & Jan 2020      \\
		CNCF Best Practices & Signing event (standard)   & May 2021      \\
		SLSA Framework      & Signing event (standard)   & Jun 2021      \\
		\bottomrule
	\end{tabular}
	\label{tab:registry_changes}
\end{table}

\iffalse
	Since we are using a difference-in-differences approach to test our signing adoption theory, we compare the adoption of signatures across registries over time.
	To do this, we measure the adoption of signatures between registries around times of change.
	We then observe the differences in changes between registries to test our hypotheses.

	Three key changes have been made to the selected registries over time.
	First, \pypi de-emphasized the use of PGP on \pypiPGPDeemp.
	Second, \pypi removed the ability to upload PGP signatures on \pypiPGPRemoved.
	Third, Cosign added support for \docker on \dockerCosignAdded.
	We summarize these changes in \cref{tab:registry_changes}.

	\hf and \maven have not experienced any changes to their signing infrastructure or policies during the time period of this study.
	For this reason, these two registries are used as control groups and use them to measure relative change in signing adoption in the other registries.
	Furthermore, we use \pypi as a control group for \docker and \docker as a control group for \pypi.
	Since we observe three separate changes, we can perform three separate comparisons.
\fi

% -------------------------------------------------------------------------------------------------------------------
% -------------------------------------------------------------------------------------------------------------------
% -------------------------------------------------------------------------------------------------------------------
\section{Results} \label{sec:results}

%In this section, we present the results of our study.
For RQ1, we present the quantity and quality of signatures we measured in each registry in~\cref{sec:results_quantity_quality}.
For RQ2, we describe changes in signing practices over time in~\cref{sec:results_change_over_time}.
Finally, for RQ3, we assess the influence of incentives on signing adoption in~\cref{sec:results_incentive}.

\subsection{RQ1: Quantity and Quality of Signatures} \label{sec:results_quantity_quality}

{
	%\arraystretch{1.5}
	\begin{table*}[!t]
		\centering
		\caption{
			The number of assessed packages and artifacts from each registry, the percent with and without signatures, and the breakdown of signature status for signed artifacts.
			For each measurement, we show the most recent year and the entire measurement period.
			``---'': Not measurable; Hugging Face hides keys, only disclosing whether validation succeeded.
		}
		\normalsize
		\begin{tabular}{lcccccc}
			\toprule
			\textbf{Registry}           & \textbf{\pypi}                                                      & \textbf{\maven}                                                       & \textbf{\docker}                                                        & \textbf{\hf}                                                                      \\
			                            & 1 year (Total)                                                      & 1 year (Total)                                                        & 1 year (Total)                                                          & 1 year (Total)                                                                    \\
			\midrule
			Total Packages              & \summaryyr[pypi][num_packages_h] (\summary[pypi][num_packages_h])   & \summaryyr[maven][num_packages_h] (\summary[maven][num_packages_h])   & \summaryyr[docker][num_packages_h] (\summary[docker][num_packages_h])   & \summaryyr[huggingface][num_packages_h] (\summary[huggingface][num_packages_h])   \\
			Total Versions              & \summaryyr[pypi][num_versions_h] (\summary[pypi][num_versions_h])   & \summaryyr[maven][num_versions_h] (\summary[maven][num_versions_h])   & \summaryyr[docker][num_versions_h] (\summary[docker][num_versions_h])   & \summaryyr[huggingface][num_versions_h] (\summary[huggingface][num_versions_h])   \\
			Total Artifacts             & \summaryyr[pypi][num_artifacts_h] (\summary[pypi][num_artifacts_h]) & \summaryyr[maven][num_artifacts_h] (\summary[maven][num_artifacts_h]) & \summaryyr[docker][num_artifacts_h] (\summary[docker][num_artifacts_h]) & \summaryyr[huggingface][num_artifacts_h] (\summary[huggingface][num_artifacts_h]) \\ \midrule
			\textbf{Unsigned Artifacts} & \summaryyr[pypi][no_sig_p]  (\summary[pypi][no_sig_p])              & \summaryyr[maven][no_sig_p]  (\summary[maven][no_sig_p])              & \summaryyr[docker][no_sig_p] (\summary[docker][no_sig_p])               & \summaryyr[huggingface][no_sig_p]  (\summary[huggingface][no_sig_p])              \\
			\textbf{Signed Artifacts}   & \summaryyr[pypi][signed_p]  (\summary[pypi][signed_p])              & \summaryyr[maven][signed_p]  (\summary[maven][signed_p])              & \summaryyr[docker][signed_p] (\summary[docker][signed_p])               & \summaryyr[huggingface][signed_p]  (\summary[huggingface][signed_p])              \\ \hline
			Good Signature              & \summaryyr[pypi][good_p]    (\summary[pypi][good_p])                & \summaryyr[maven][good_p]    (\summary[maven][good_p])                & 100\% (100\%)                                                           & \summaryyr[huggingface][good_p]    (\summary[huggingface][good_p])                \\
			Bad Signature               & \summaryyr[pypi][bad_sig_p] (\summary[pypi][bad_sig_p])             & \summaryyr[maven][bad_sig_p] (\summary[maven][bad_sig_p])             & 0.0\% (0.0\%)                                                           & ---                                                                               \\
			Expired Signature           & \summaryyr[pypi][exp_sig_p] (\summary[pypi][exp_sig_p])             & \summaryyr[maven][exp_sig_p] (\summary[maven][exp_sig_p])             & 0.0\% (0.0\%)                                                           & ---                                                                               \\
			Expired Public Key          & \summaryyr[pypi][exp_pub_p] (\summary[pypi][exp_pub_p])             & \summaryyr[maven][exp_pub_p] (\summary[maven][exp_pub_p])             & 0.0\% (0.0\%)                                                           & ---                                                                               \\
			Missing Public Key          & \summaryyr[pypi][no_pub_p]  (\summary[pypi][no_pub_p])              & \summaryyr[maven][no_pub_p]  (\summary[maven][no_pub_p])              & 0.0\% (0.0\%)                                                           & ---                                                                               \\
			Public Key Revoked          & \summaryyr[pypi][rev_pub_p] (\summary[pypi][rev_pub_p])             & \summaryyr[maven][rev_pub_p] (\summary[maven][rev_pub_p])             & 0.0\% (0.0\%)                                                           & ---                                                                               \\
			Bad Public Key              & \summaryyr[pypi][bad_pub_p] (\summary[pypi][bad_pub_p])             & \summaryyr[maven][bad_pub_p] (\summary[maven][bad_pub_p])             & 0.0\% (0.0\%)                                                           & ---                                                                               \\
			\bottomrule
		\end{tabular}
		\label{tab:raw-numbers}
	\end{table*}
}

In \cref{tab:raw-numbers}, we show the quantity and quality of signatures in each registry.
We show the total amount of signable artifacts in each registry, how many of those are signed, and the status of the subset of signed signatures.
We show both the most recent year of data (Jan-Dec 2023) and the entire time period (\startDate to \finishDate).

\subsubsection{Quantity of Artifact Signatures}
With respect to the quantity (proportion) of signed artifacts, the registries lie in three groups by order of magnitude.
First, \textit{\maven} experiences the highest signing rate with \summaryyr[maven][signed_p] of artifacts signed in 2023.
We conjecture that this degree of signing occurs only when signing is mandatory.
\footnote{
	Not all \maven packages are signed.
	Some are ingested from other Java package registries and the Maven signing requirement is waived.
	\textit{Source: Personal communication with Maven team.}
}
Second, \textit{\docker} has a low adoption rate with \summaryyr[docker][signed_p] of tags signed in 2023.
Third, \textit{\hf} and \textit{\pypi} currently have a negligible amount of signatures with \summaryyr[huggingface][signed_p] and \summaryyr[pypi][signed_p] of artifacts signed in 2023, respectively.
Keep in mind that \textit{\pypi's} low signing rate includes data since the feature was removed in \pypiPGPRemoved.
Even considering this, the relative number of signed artifacts is still the lowest on \textit{\hf}.
Out of all \summaryyr[huggingface][num_artifacts_h] commits across the \summaryyr[huggingface][num_packages_h] packages on \textit{\hf} in 2023, only \summaryyr[huggingface][signed_h] are signed.

\begin{tcolorbox} [width=\linewidth, colback=yellow!30!white, top=1pt, bottom=1pt, left=2pt, right=2pt]
	\textbf{Finding 1}:
	Between \startYear and \finishDate, all registries aside from \maven had less than 2\% of artifacts signed.
	\maven, the only registry in our study that mandates signing, had \summaryyr[maven][signed_p] of artifacts signed in that same time period.
\end{tcolorbox}

\subsubsection{Quality of Artifact Signatures}

\myparagraph{Failure Modes}
With respect to quality, each registry has a distinct flavor.
On \textit{\docker} signatures either exist or not --- we can only tell if maintainers correctly signed a tag.
On \textit{\hf}, we can only tell if the signature was valid.
On \textit{\maven} and \textit{\pypi}, we can measure the failure modes of signatures.

% \JD{MUSTFIX Seems like similar remarks are needed for maven HF and pypi? Then after that, start a new paragraph.}
Of the registries with measurable quality, \textit{\maven} has the best with \summaryyr[maven][good_p] of signatures valid since \startYear.
\textit{\pypi} has the next best quality with \summaryyr[pypi][good_p] of signatures valid since \startYear.
Not only does \textit{\hf} have the lowest quantity of signed artifacts, but it also has the lowest quality of signed artifacts.
Of the \summaryyr[huggingface][signed_h] signed artifacts, only \summaryyr[huggingface][good_h] (\summaryyr[huggingface][good_p]) were valid.

We observed differences between signing failure modes by registry.
The three most common failure modes on \textit{\maven} were expired public keys, missing public keys, and bad public keys.
Failures related to public keys accounted for over 99\% of all Maven signing failures in 2023. %\startYear and \finishDate.
The three most common failure modes on \textit{\pypi} were missing public keys, revoked public keys, and expired public keys.
Expired signatures are very rare, the only instances we could find were from 2014 versions of the leekspin package on \pypi
Similar to \maven, public key related failures accounted for over 99\% of all PyPI signing failures in 2023. %between \startYear and \finishDate.
On \hf, the registry records but does not publish the public keys disclosed by package maintainers.\footnote{We asked the Hugging Face engineers for access. They declined.}
Due to the lack of published public keys, we were unable to determine the cause of the invalid signatures.
Finally, we observed no signing failures in Docker Hub.

\begin{tcolorbox} [width=\linewidth, colback=yellow!30!white, top=1pt, bottom=1pt, left=2pt, right=2pt]
	\textbf{Finding 2}:
	Signing failures were common in three of the four studied registries.
	On \maven and \pypi, 24.0\% and 53.1\% of signatures between \startYear and \finishDate were invalid, respectively.
	On \hf the situation is worse, with \summaryyr[huggingface][bad_sig_p] invalid signatures.
	Lastly, on \docker, we observed no signing failures.
\end{tcolorbox}

{
\begin{table}
	\centering
	\caption{
		Cryptographic algorithms used in signatures.
	}
	% \normalsize
	\begin{tabular}
		{lcc}
		\toprule
		\textbf{Algorithm} & \textbf{\maven}                       & \textbf{\pypi}                       \\
		\toprule
		\textbf{RSA}       & \textbf{\crypto[maven][rsa][percent]} & \textbf{\crypto[pypi][rsa][percent]} \\
		\textbf{DSA}       & \crypto[maven][dsa][percent]          & \textbf{\crypto[pypi][dsa][percent]} \\
		EdDSALegacy        & \crypto[maven][eddsa][percent]        & \crypto[pypi][eddsa][percent]        \\
		RSA Sign Only      & \crypto[maven][rsasignonly][percent]  & \crypto[pypi][rsasignonly][percent]  \\
		ECDSA              & \crypto[maven][ecdsa][percent]        & \crypto[pypi][ecdsa][percent]        \\
		\bottomrule
	\end{tabular}
	\label{tab:crypto}
\end{table}
}

{
\begin{table}
	\centering
	\caption{
		RSA key lengths used in signatures.
	}
	% \normalsize
	\begin{tabular}
		{ccc}
		\toprule
		\textbf{Length} & \textbf{\maven}                                        & \textbf{\pypi}                                        \\
		\toprule
		8192            & \crypto[maven][rsa][key_sizes][8192][percent]          & \crypto[pypi][rsa][key_sizes][8192][percent]          \\
		4608            & \crypto[maven][rsa][key_sizes][4608][percent]          & \crypto[pypi][rsa][key_sizes][4608][percent]          \\
		\textbf{4096}   & \textbf{\crypto[maven][rsa][key_sizes][4096][percent]} & \textbf{\crypto[pypi][rsa][key_sizes][4096][percent]} \\
		\textbf{3072}   & \textbf{\crypto[maven][rsa][key_sizes][3072][percent]} & \crypto[pypi][rsa][key_sizes][3072][percent]          \\
		\textbf{2048}   & \textbf{\crypto[maven][rsa][key_sizes][2048][percent]} & \textbf{\crypto[pypi][rsa][key_sizes][2048][percent]} \\
		1536            & \crypto[maven][rsa][key_sizes][1536][percent]          & \crypto[pypi][rsa][key_sizes][1536][percent]          \\
		1024            & \crypto[maven][rsa][key_sizes][1024][percent]          & \crypto[pypi][rsa][key_sizes][1024][percent]          \\
		\bottomrule
	\end{tabular}
	\label{tab:rsa_length}
\end{table}
}

\myparagraph{Cryptographic Algorithms}
For \maven and \pypi, we observed the use of several cryptographic algorithms.
We show the distribution of algorithms in~\cref{tab:crypto}.
RSA was the most common algorithm used in both \maven and \pypi.
In \maven, RSA was used in \crypto[maven][rsa][percent] of signatures.
In \pypi, RSA was used in \crypto[pypi][rsa][percent] of signatures.

RSA signature security is dependent on the key length.
In~\cref{tab:rsa_length}, we show the distribution of RSA key lengths used in \maven and \pypi.
Of the RSA signatures in \maven, most of them used either 2048 or 4096 bit keys.
The same is true for \pypi.
These keysizes comply with the US NIST's SP-800-78-5 baseline~\cite{NIST_RSA_SIZES} for keysizes for this decade.
Only a small fraction of signatures used keys larger than 4096 bits or smaller than 2048 bits.

\begin{tcolorbox} [width=\linewidth, colback=yellow!30!white, top=1pt, bottom=1pt, left=2pt, right=2pt]
	\textbf{Finding 3}:
	For both \maven and \pypi, RSA was the most common cryptographic algorithm used in signatures.
	Most RSA signatures used 2048 or 4096 bit keys.
	A small fraction used insecurely-small key sizes ($<$2048 bits) or very large key sizes ($>$4096 bits).
\end{tcolorbox}

\myparagraph{Expired Keys}
We report expired keys as a failure mode regardless of the publishing time of the artifact.
For old artifacts, this may be unfair, since the key was presumably valid at time of publication, and since a newer version of the package may have been available.
We investigated both of these aspects for signatures whose keys had expired.

First, we examined the typical time of validity, \ie the time remaining in the public key's lifespan at time of signature creation.
Surprisingly, a substantial proportion of signatures are created \textbf{after} the expiration of the associated public key, \ie it was never a valid signature.
For \maven, \expkeys[maven][pub_exp_before][percent] of the artifacts whose public key eventually expired had signatures created after the expiration.
For \pypi, \expkeys[pypi][pub_exp_before][percent] of the artifacts whose public key eventually expired had signatures created after the expiration.
For signatures created with a still-valid public key,
signatures on  \maven had a median of \expkeys[maven][median_remaining] years remaining in the public key's lifespan and
those on \pypi had a median of \expkeys[pypi][median_remaining] years remaining in the public key's lifespan.

Second, we examined the availability of an upgrade path from an expired to an unexpired version of a package.
%For package versions with expired public keys, we found that only some of them could upgrade to valid signed versions after the key expiration.
On \maven, in only \expkeys[maven][upgrade_paths][upgradable][percent] of cases was there a newer version of the package with an unexpired signature. %  packages with expired keys could upgrade to a signed version, the rest had no easy upgrade path.
On \pypi, the number was \expkeys[pypi][upgrade_paths][upgradable][percent].
Thus, upgrade paths are usually unavailable, suggesting that signature expiration is not well managed in these ecosystems.
This result may be confounded by abandoned packages. %, which are difficult to distinguish.

\subsection{RQ2: Change in Signing Practices Over Time} \label{sec:results_change_over_time}

\subsubsection{Quantity of Artifact Signatures}

\begin{figure}
	\centering
	\includegraphics[width=\linewidth]{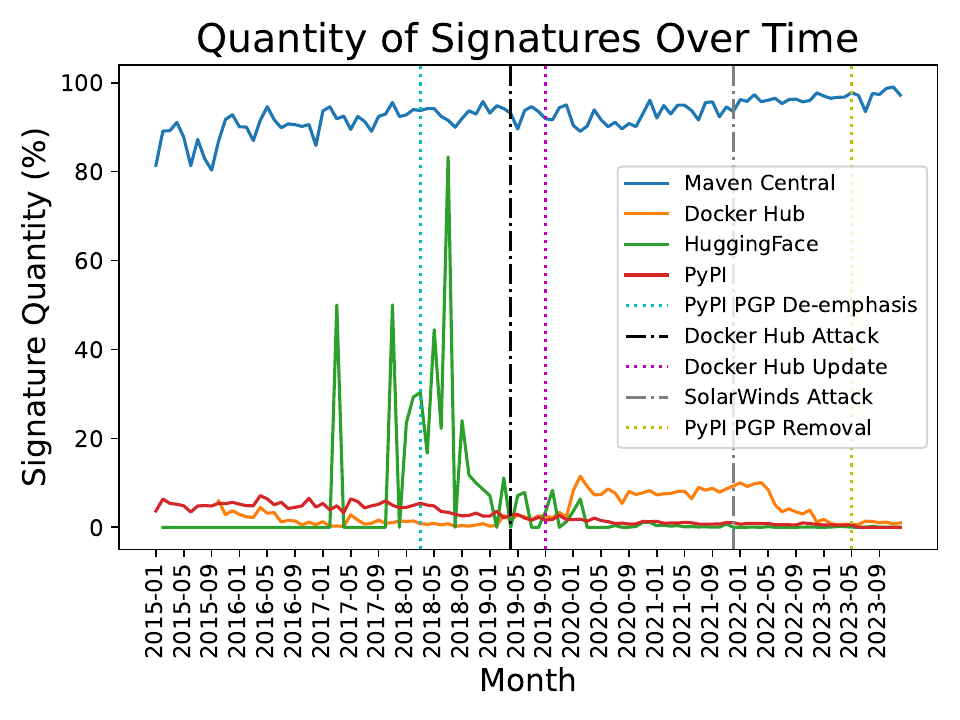}
	\caption{
		Quantity of signed artifacts over time.
		Axes are time (monthly increments)
		and the percentage of signed artifacts per registry.
	}
	\label{fig:results_quantity_combined}
\end{figure}

In \cref{fig:results_quantity_combined}, we show the quantity of signed artifacts over time.
We measured signing rates for the signable artifacts published in that month.
This figure shows how many such artifacts were signed, grouped by registry.

We observe a stark contrast in signing quantity between \maven and the other registries.
Because of its mandatory signing policy as of 2005, \maven had a high quantity of signed artifacts throughout the period we measured.
In contrast, the other registries have had a low quantity of signed artifacts throughout.
\pypi, for example, has had a historically decreasing quantity of signed artifacts until they were ultimately removed in \pypiPGPRemoved.
\hf has also experienced a low signing rate across its lifespan.
\footnote{
	The \hf spikes in~\cref{fig:results_quantity_combined} are due to the small number of commits in Jan 2017--Dec 2019 (only 1,266 commits in this period).
	%Only 1,266 commits were created in Jan 2017--Dec 2019.
}
%Even with its recent increase in popularity, signing has not been adopted by the community.
Before late 2019, \docker had a worse signing rate than \pypi.
From early 2020 through the middle of 2022, Docker experienced a notable increase in signing.

\begin{tcolorbox} [width=\linewidth, colback=yellow!30!white, top=1pt, bottom=1pt, left=2pt, right=2pt]
	\textbf{Finding 4}:
	\maven is the only registry with a consistently high quantity of signed artifacts.
\end{tcolorbox}

\subsubsection{Quality of Artifact Signatures}

\begin{figure}
	\centering
	\includegraphics[width=\linewidth]{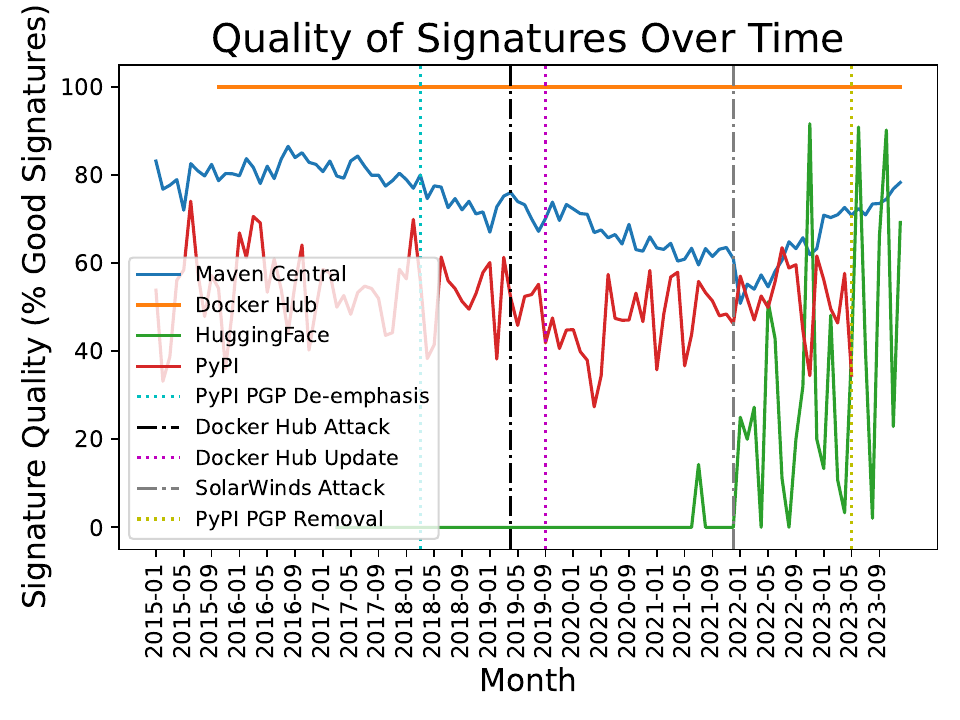}
	\caption{
		Quality of signed artifacts over time.
		X-axis shows time (monthly increments).
		Y-axis shows percentage of signatures with good status.
	}
	\label{fig:results_quality_combined}
\end{figure}

\myparagraph{Failure Modes}
In \cref{fig:results_quality_combined}, we show the quality of signed artifacts over time.
This figure shows how many of the signed artifacts in each registry were signed correctly in a given month.
Within each registry, we observe no change in the quality over time.
Between registries, we observe perfect quality in Docker Hub; high quality in Maven, lower and variable quality in PyPI, and spikes (due to the low number of signatures) of quality in Hugging Face.

\begin{figure}
	\centering
	\includegraphics[width=\linewidth]{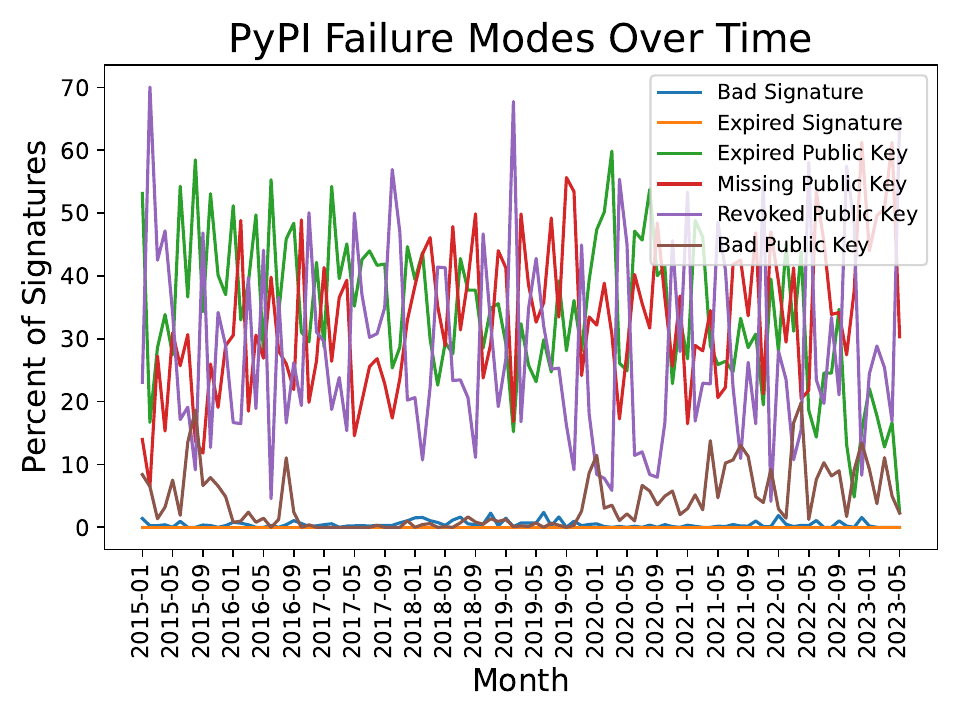}
	\caption{Failure modes of signatures on PyPI.}
	\label{fig:results_failures_pypi}
\end{figure}

\begin{figure}
	\centering
	\includegraphics[width=\linewidth]{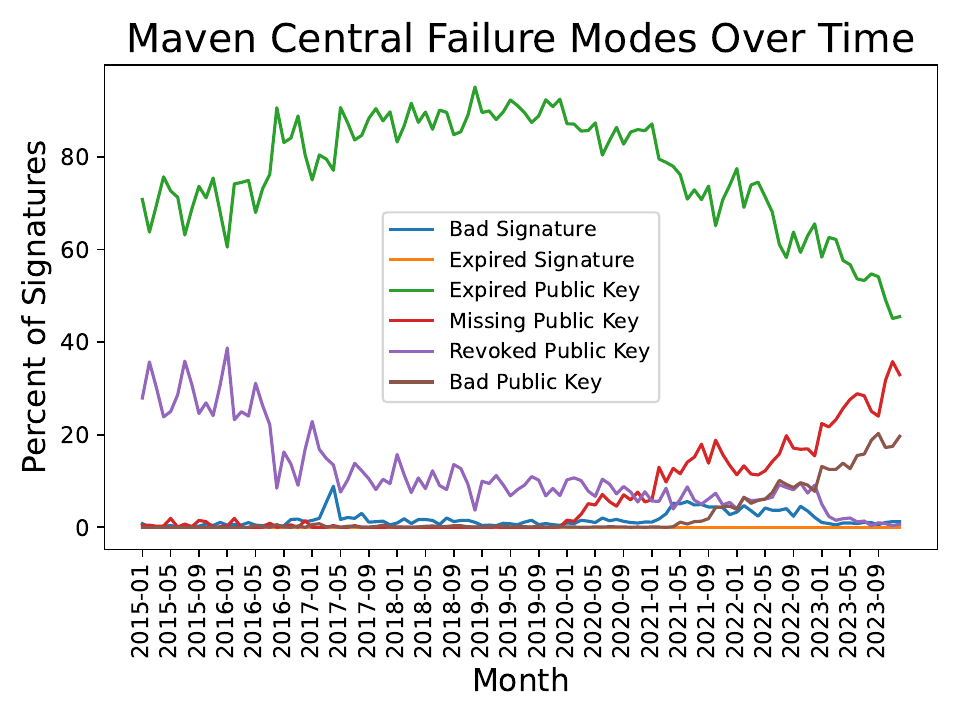}
	\caption{Failure modes of signatures on Maven Central.}
	\label{fig:results_failures_maven}
\end{figure}

Next we consider the failure modes of signatures by registry.
For \textit{\pypi}, see~\cref{fig:results_failures_pypi}. %, we show the failure modes of signatures on \pypi.
There are several common failure modes.
They vary in relative frequency and none dominates.
For signing failure modes in \textit{\maven} see~\cref{fig:results_failures_maven}. %, we show the failure modes of signatures on \maven.
The primary failure mode in our study period is an expired public key.
Revoked public keys have become less of a concern over time and missing public keys have become more of a concern since the end of 2019.
Bad public keys are also on the rise.
Public key creation and distribution remains challenging. %maintainers are getting worse at public key creation and distribution.

We omit figures for \docker and \hf.
Since \docker signatures are either valid or invalid (and all we measured were valid), we cannot distinguish the failure modes. %do not show the failure modes of signatures on \docker.
On \hf, we cannot access the maintainers' PGP keys, so we cannot analyze the failure modes of signatures there.

\begin{tcolorbox} [width=\linewidth, colback=yellow!30!white, top=1pt, bottom=1pt, left=2pt, right=2pt]
	\textbf{Finding 5}:
	\docker is the only registry with perfect quality.
	For \maven and \pypi, the most common failure modes are related to public keys in our study period.
\end{tcolorbox}
% \JD{MUSTFIX: This box appears twice --- it also appeared on the previous page.}

\begin{figure}
	\centering
	\includegraphics[width=\linewidth]{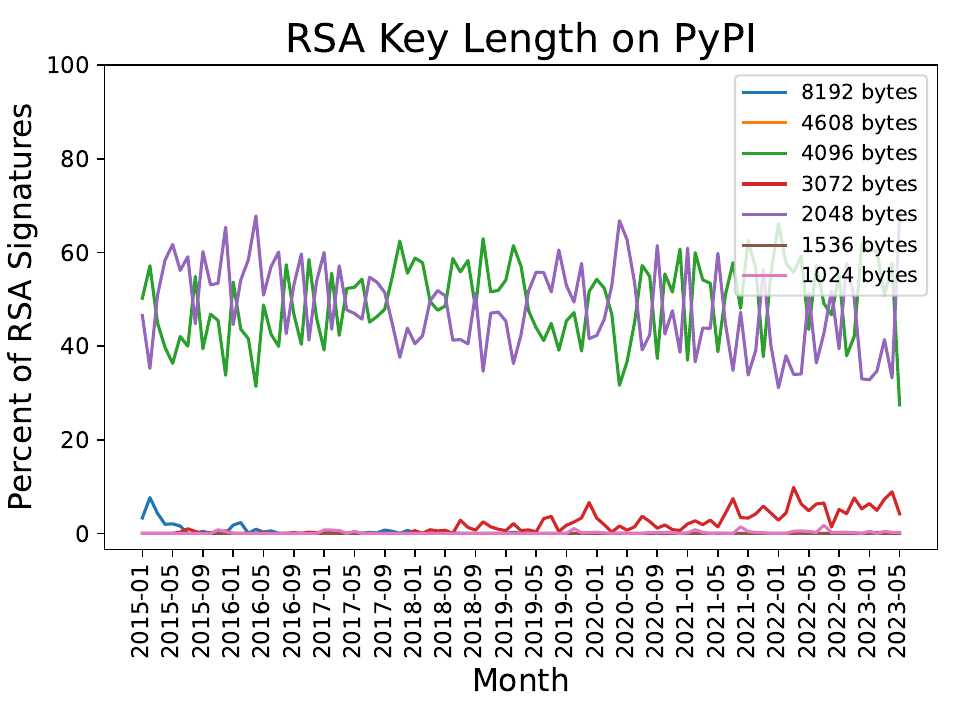}
	\caption{RSA key length over time in \pypi.}
	\label{fig:rsa_pypi}
\end{figure}

\myparagraph{Cryptographic Algorithms}
For RSA keys on \pypi,~\cref{fig:rsa_pypi} shows the key lengths used over time.
Note that 2048 and 4096 bit keys trade off as the most common over time.
3072 bit keys are also used starting in mid-2018, but remain much less common than the other two key lengths.

\begin{figure}
	\centering
	\includegraphics[width=\linewidth]{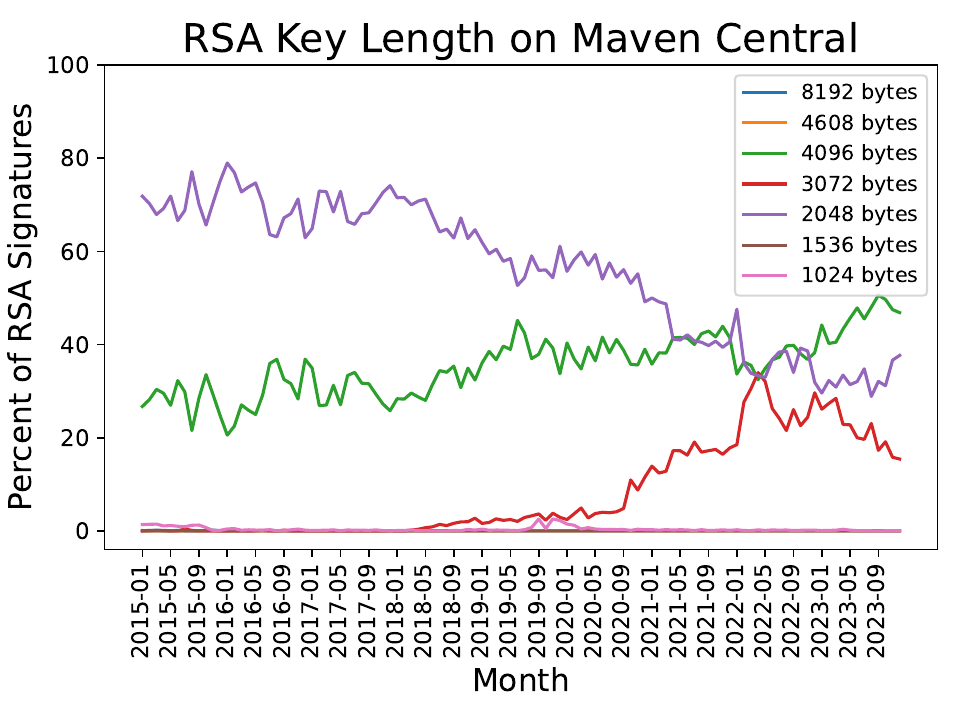}
	\caption{RSA key length over time in \maven.}
	\label{fig:rsa_maven}
\end{figure}

For RSA keys on \maven,~\cref{fig:rsa_maven} shows the key lengths used over time.
2048 bit keys are initially the most common, but then drop to a similar level as the 4096 bit keys.
As in \pypi, 3072 bit keys start to be used in mid-2018.
3072 bit keys start to replace 2048 bit keys in late 2020 but are still less common than 2048 and 4096 bit keys.

\begin{tcolorbox} [width=\linewidth, colback=yellow!30!white, top=1pt, bottom=1pt, left=2pt, right=2pt]
	\textbf{Finding 6}:
	2048 and 4096 bit RSA keys have remained the most common key lengths in both \maven and \pypi between \startDate and \finishDate.
	On \maven, 3072 bit keys started to replace 2048 bit keys in late 2020.
\end{tcolorbox}

\subsubsection{No bias from new packages}

\begin{figure}
	\centering
	\includegraphics[width=\linewidth]{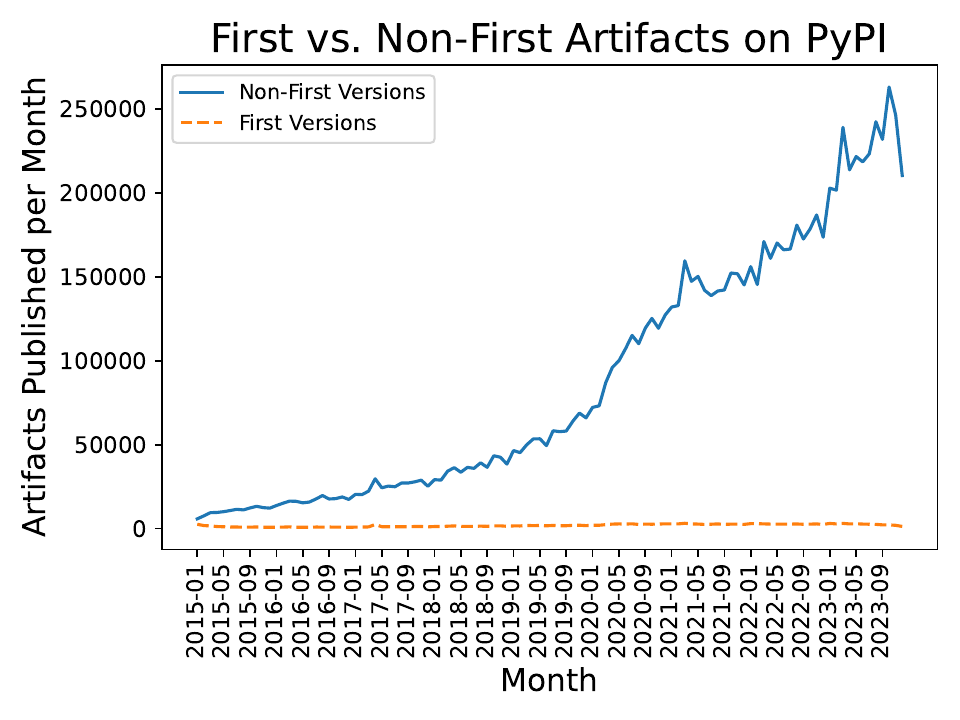}
	\caption{
 Number of first-artifacts and subsequent-artifacts of packages on PyPI.}
	\label{fig:new_artifacts_pypi}
\end{figure}

Software package registries grow over time~\cite{ecosystems_website}.
One consideration about our longitudinal analysis is therefore whether the signing practices of new packages dominates our measure in each time window.
To assess this possibility,~\cref{fig:new_artifacts_pypi} shows the number of first-versions (\ie a new package) and subsequent-versions (\ie a new version of an existing package) of packages on \pypi, binned monthly.
%We considered that some of our results may have been skewed by an increase in new packages since they be less likely to have signatures.
We note that most of the artifacts on \pypi are from subsequent-versions of packages.
\maven, \docker, and \hf follow the same trend.
Hence, our results reflect the ongoing practice of existing maintainers rather than the recurring mistakes of new maintainers.

\iffalse
	Most of the artifacts published each month are from subsequent-versions of packages.
	This suggests that the lack of signing adoption is not due to an influx of new packages.
\fi

\subsection{RQ3: Influence of Incentive} \label{sec:results_incentive}

To test our hypotheses, we performed a one-way ANOVA test, subsequent Tukey tests, and a selection of one-sided, two-way t-tests.
The dependent variable is signature quantity, \ie daily signing rate (percent of signed artifacts).

For the one-way ANOVA test, we compared the signing rates over the entire sample period between each of our registries.
We received a p-value of \stats[anova][p] and an F-statistic of \stats[anova][f].
This indicates a significant difference in signing rates between the registries.

To determine which registries are different, we performed a Tukey test.
The Tukey test performs pairwise comparisons between the registries.
In all cases except for the comparison between \pypi and \hf (p-value of \stats[tukey][pypi][huggingface][p]), we observed a significant difference in signing rates (all p-values of \stats[tukey][pypi][maven][p]).
This indicates that the signing rates of all registries are significantly different from each other except for \pypi and \hf.

We use these results, and a series of one-sided, two-way t-tests to evaluate our hypotheses.
The summary of those t-tests is shown in~\cref{tab:ttests}.
Note that we do not include \hf in our t-tests because it has a negligible quantity of signed artifacts during the time periods surrounding the selected events.

\begin{table}
	\centering
	\caption{
		Results of t-tests for each hypothesis.
		Per event, we show the registry, adjusted p-value, and effect size.\\
        *: Statistically significant value at p $<$ 0.01.
	}
	\begin{tabular}{llcc}
		\toprule
		\textbf{Event}   & \textbf{Registry} & \textbf{Adj. P-Val}                         & \textbf{Effect Size} \\
		\midrule
		PyPI De-emph     & PyPI              & \textbf{\stats[ttests][pypi_deemp][pypi][adj_p]*}      & \stats[ttests][pypi_deemp][pypi][effect_size]               \\
		PyPI De-emph     & Maven Central     & \stats[ttests][pypi_deemp][maven][adj_p]     & \stats[ttests][pypi_deemp][maven][effect_size]              \\
		PyPI De-emph     & Docker Hub        & \textbf{\stats[ttests][pypi_deemp][docker][adj_p]*}    & \stats[ttests][pypi_deemp][docker][effect_size]             \\
		\midrule                                                                                                                        
		DockerHub Update & PyPI              & \stats[ttests][docker_update][pypi][adj_p]   & \stats[ttests][docker_update][pypi][effect_size]            \\
		DockerHub Update & Maven Central     & \stats[ttests][docker_update][maven][adj_p]  & \stats[ttests][docker_update][maven][effect_size]           \\
		DockerHub Update & Docker Hub        & \textbf{\stats[ttests][docker_update][docker][adj_p]*} & \stats[ttests][docker_update][docker][effect_size]          \\
		\midrule                                                                                                                        
		DockerHub Hack   & PyPI              & \stats[ttests][docker_hack][pypi][adj_p]     & \stats[ttests][docker_hack][pypi][effect_size]              \\
		DockerHub Hack   & Maven Central     & \stats[ttests][docker_hack][maven][adj_p]    & \stats[ttests][docker_hack][maven][effect_size]             \\
		DockerHub Hack   & Docker Hub        & \stats[ttests][docker_hack][docker][adj_p]   & \stats[ttests][docker_hack][docker][effect_size]            \\
		\midrule                                                                                                                        
		SolarWinds       & PyPI              & \stats[ttests][solarwinds][pypi][adj_p]      & \stats[ttests][solarwinds][pypi][effect_size]               \\
		SolarWinds       & Maven Central     & \stats[ttests][solarwinds][maven][adj_p]     & \stats[ttests][solarwinds][maven][effect_size]              \\
		SolarWinds       & Docker Hub        & \stats[ttests][solarwinds][docker][adj_p]    & \stats[ttests][solarwinds][docker][effect_size]             \\
		\bottomrule
	\end{tabular}
	\label{tab:ttests}
\end{table}

\subsubsection{H$_1$: Registry Policies}
H$_1$ predicts that registry policies will either encourage or discourage signing adoption quantity.
If this is the case, we would expect to see corresponding changes in the measurements shown in \cref{fig:results_quantity_combined}.
For \maven and \hf, we did not identify any changes in registry policies during the time period of this study.
We did, however, identify two policy changes in \pypi and one change in \docker.
These are shown in~\cref{tab:registry_changes} and appear as vertical lines in~\cref{fig:results_quantity_combined}.

Three observations from this figure support hypothesis H$_1$.
%First, \pypi's quantity of signatures has steadily been decreasing over the past several years.
On \pypiPGPDeemp, \pypi de-emphasized the use of PGP by removing UI elements that encouraged signing.
%Shortly after this, the quantity of signatures on \pypi began to decrease.
%This continued until \pypiPGPRemoved, when \pypi removed the ability to upload PGP signatures entirely.
%After that point, the quantity of signatures on \pypi dropped to zero.
\textit{Our t-tests show a statistically significant change (decrease) in signing rates on PyPI after this event.
}We also measured an effect on \docker, although the effect size is small and referring to~\cref{fig:results_quantity_combined} there is no clear long-term trend.

%that the only registries with a statistically significant change in signing rate after the de-emphasis of PGP were \pypi and \docker.
%The smaller effect size of \docker suggests that the de-emphasis of PGP had a larger effect on \pypi than \docker.

Second, \docker's quantity of signatures experienced an appreciable increase at the start of 2020.
This increase follows a provenance visibility related update to Docker Hub, published on 05 Sep 2019.\footnote{See \url{https://docs.docker.com/docker-hub/release-notes/\#2019-09-05}.}
In this update, Docker Hub increased tag visibility and updated security scan summaries.
This update may have encouraged maintainers to sign their tags.
\textit{Our t-tests show that the only registry with a statistically significant change in signing rate after the Docker Hub update was \docker.}

Third, the notable difference in signing quantity between \maven and the other registries suggests that mandatory signing policies encourage adoption.
Since \maven has a mandatory signing policy, we expected, and observe, a high quantity of signatures.
\textit{Our ANOVA test, discussed earlier, found that the signing rate of \maven is significantly different (higher) than the others.} % registries.}

\subsubsection{H$_2$: Dedicated Tooling}

H$_2$ predicts that dedicated tooling will affect both the quantity and quality of signatures.
Since \docker is the only registry in our study that has dedicated tooling, H$_2$ predicts that \docker would have a higher quantity and quality of signatures than the other registries.

We do not observe support for the quantity aspect of H$_2$.
In \cref{fig:results_quantity_combined}, we observe that \docker has a lower quantity of signatures than \maven and had lower quantity of signatures than \pypi before 2020.
This suggests that the dedicated tooling on \docker did not significantly impact the quantity of signatures.
After all, between \startYear and \finishDate, \docker's signing rate was only \summaryyr[docker][signed_p].

However, we do find support for the quality aspect of H$_2$.
In \cref{fig:results_quality_combined}, we observe that \docker has perfect signature quality --- something no other registry can achieve.
This is because all signatures on \docker must be created with through the DCT which, in itself, checks to make sure signatures are created correctly.

\subsubsection{H$_3$: Cybersecurity Events}

H$_3$ predicts that cybersecurity events will encourage signing adoption quantity and quality.
Since these events are not specific to any registry, we would expect to see corresponding changes in the measurements shown in \cref{fig:results_quantity_combined} and \cref{fig:results_quality_combined} from multiple registries.
\cref{tab:registry_changes} lists several influential software supply chain attacks and cybersecurity events.
However, neither preliminary visual analysis nor t-tests show any significant changes in signing rate or quality after these events.

We illustrate this with two cases.
First, consider the registry-specific Docker Hub hack in April 2019.
Although this attack was widely publicized and required over 100,000 users to take action to secure their accounts, this attack had little observable effect on the quantity of signatures on \docker.
The large increase in signing adoption on \docker occurred at the start of 2020, about 9 months after the Docker Hub hack (and just after a registry-specific policy change, to the visibility of package signatures).
This implies that the attack did not even have an impact on the quantity of signatures of its victim registry.
Other registries were not affected at all.

Second, the SolarWinds attack in December 2020 had little effect on the quantity of signatures for any registry.
This attack was one of the largest software supply chain attacks in history.
It led to several government initiatives to improve software supply chain security. %, even these initiatives have not seemed to have any appreciable impact on signature quantity or quality.
However, SolarWinds (and those government initiatives, \eg the subsequent executive order and NIST guidance) had no discernible effect on signing adoption in the studied registries.

\subsubsection{H$_4$: Startup Cost}

H$_4$ predicts that the first signature in a package will encourage subsequent signing.
This is relatively simple to measure.
First, we determine the probability of an artifact having a signature in each registry.
We then determine the probability of an artifact having a signature if one of the previous artifacts from the same package has been signed.
We then compare these two probabilities to determine if the first signature predicts subsequent signing.

In \cref{tab:metric}, we show both of these probabilities for each of our four registries.
All registries experience an increase from the raw probability to the probability after the first signature.
This suggests that overcoming the burden of signing for the first time encourages subsequent signing.
The magnitude of this increase varies by registry.
On \maven, we observe a small increase in signing probability, but this is expected since \maven has a mandatory signing policy.
On the other platforms, the increase was $\sim$40x.
%Both \pypi and \docker observed similar increases in probability (33\% and 31\% respectively).
%Finally, \hf observed the largest increase in probability (52\%).
These changes suggest that the initial burden of signing is a significant barrier to adoption.

\begin{table}[ht]
	\centering
	\small
	\caption{
		The probability of an artifact having a signature.
		Raw probability describes the likelihood of any artifact in the registry having a signature.
		After $1^{st}$ Signature describes the probability that an artifact will be signed if one of the previous artifacts from the same package has been signed.
	}
	\begin{tabular}{lcc}
		\toprule
		\textbf{Registry Name} & \textbf{Raw Probability}       & \textbf{After $1^{st}$ Signature} \\ \toprule
		PyPI                   & \metric[pypi][chance_p]        & \metric[pypi][metric_p]           \\
		Maven Central          & \metric[maven][chance_p]       & \metric[maven][metric_p]          \\
		Docker Hub             & \metric[docker][chance_p]      & \metric[docker][metric_p]         \\
		Hugging Face           & \metric[huggingface][chance_p] & \metric[huggingface][metric_p]    \\
		\bottomrule
	\end{tabular}
	\label{tab:metric}
\end{table}

% -------------------------------------------------------------------------------------------------------------------
% -------------------------------------------------------------------------------------------------------------------
% -------------------------------------------------------------------------------------------------------------------

\section{Discussion} \label{sec:discussion}
% TODO: This section will need to be modified once we get results

\noindent
We highlight three points for discussion.

First, our findings suggest that the long line of literature on the usability of signing tools (\cref{sec:related_works}) may benefit from extending its perspective from an individual view to ecosystem-level considerations.
Two registries, Maven and PyPI, use the same PGP-based signing method.
We observe significant variations in signing adoption between these two registries, both in signature quantity and in signature quality/failure modes.
Our answers to RQ3 suggest that the registry policies have a substantial effect on signing adoption, regardless of the available tooling.
However, we acknowledge that our data do substantiate their concern about signature quality ---
our data expose major issues with signature quality in both the Maven and the PyPI registries,
and
that the dedicated tooling available in \docker appears to eliminate the issues of signing quality.

Second, our findings suggest that registry operators control the largest incentives for software signing.
Mandating signatures has not apparently decreased the popularity of Maven --- we recommend that other registries do so.
Registry operators can also learn from the success of \docker, whose dedicated tooling results in perfect signing quality.
No registry currently mandates signatures \textit{and} provides dedicated tooling.
Our results predict that the combination would result in high signature quantity and quality.

Third, we were disturbed at the non-impact of signing events --- software supply chain attacks, government orders, and industry standards.
Good engineering practice (not to mention engineering codes of ethics) calls for engineers to recognize and respond to known failure modes.
Our contrary results motivate continued research into engineering ethics and a failure-aware software development lifecycle~\cite{anandayuvaraj2022reflecting}.

\section{Threats to Validity}

We distinguish three kinds of threats: construct, internal, and external.

\myparagraph{Construct}
Our study operationalized several constructs.
We defined signature adoption in terms of quantity (proportion) and quality (frequency of no failure).
We believe our notion of signature quantity is unobjectionable.
However, signature quality is somewhat subjective.
We made four assumptions that may bias our results:
(1) We defined quality based on failure modes derived from the error cases of GPG;
(2) We considered expired and revoked keys as failures even if the keys were valid at the time of signing;
(3) We reported the cryptographic algorithm and key size for signatures on \pypi and \maven but did not include this as a factor in quality; and
(4) We relied on the correctness of specific (albeit widely used) tools to measure the signatures.

We acknowledge that the limitations of our data sources may potentially impact the robustness of our results.
Our reliance on third party metadata (\ie ecosyste.ms and BigQuery) may introduce errors and our key discovery methodology may fail to find keys that exist on small websites or that have been shared through other means.
In addition, our insights into failure modes were limited by the data made available by the signing infrastructure of our target registries.

\myparagraph{Internal}
We evaluated a theory of incentive-based software signing adoption based on several hypotheses.
Due to the difficulty of conducting controlled experiments of this theory, we used a quasi-experiment, and assumed no uncontrolled confounding factors.
Among \maven, \pypi, and \docker, we have no reason to believe there would be such factors.
In \hf, there may be confounding factors related to the nature of the platform itself.
As noted by Jiang \etal, \hf is characterized by a ``research to practice pipeline'' more than traditional software package registries are~\cite{jiang2022PTMSupplyChain}.
Researchers have little incentive to follow secure engineering practices.
This difference could comprise an uncontrolled confounding factor.
However, \hf had little variation in signing quantity and none of our main results relied on \hf phenomena.

\myparagraph{External}
All empirical studies are limited in generalizability by the subjects they study.
Our work examines four of the most popular software package registries, across three kinds of packages (traditional software, Docker containers, and machine learning models).
Our results thus have some generalizability.
However, our results may not generalize to contexts with substantially different properties, \eg registries more influenced by government policy or more dominated by individual organizations.

\section{Future Work}

\noindent
We suggest several directions for future research.

\myparagraph{Further Diversification of Registries}
% While Our study have mainly considered  OSS ecosystems, We propose that future works consider other types of ecosystems\eg commercial and proprietary ecosystems. We also propose that other ecosystems where other forms of signing not PGP/GPG are implemented be considered. A cross-ecosystem analysis that includes various repository types, such as open (or community), curated, and closed  repositories or even app markets is also suggested.
Our findings are provocative, but we recommend diversifying the types of ecosystems under study.
Further work could go beyond open-source registries to include commercial and proprietary registries (\eg app stores), and ecosystems implementing different forms of signing solutions.
%we propose a cross-ecosystem analysis that encompasses various repository types, such as open (or community), curated, and closed repositories, along with app markets.
This approach will facilitate a comprehensive exploration of other factors, incentives and cost trade-offs that influence the adoption of software signing for these types of ecosystems.
% \begin{itemize}
%   \item Extend the work to include more ecosystems  {larger dataset/ different ecosystems }
%   \item Use cross ecosystems types approach (current:: open or community repos /suggested: current + curated repositories/closed repos or app markets) 
%   \begin{itemize}
%    \item  the incentive models are different, as well as the meaning of a signature/the process to transmit trust 
%     \item Trust in curated repositories is vested on the curator, rather than the author, and do not transmit any information about the curator's due diligence
%     \item paid marketplaces do due diligence on the identity of the submitter, but the incentive of the marketplace is to maintain a baseline of trust in the platform and is economic in nature

\myparagraph{Incorporating Human Factors}
Our approach is grounded in a theory of software signing based in incentives.
Qualitative data --- \eg surveys and interviews of engineers in the registries of interest --- would shed light on the relative weight of the factors we identified.
Such studies could expose new factors for quantitative evaluation.

%     \end{itemize}
%   \item  Other incentive types:Monetary to sign pkg,Community based feedback ,Monitoring incentives
%   \item Have more systematic theories of incentives 
%   \begin{itemize}
%       \item  Interviewing oss developer ( why do you sign ? What are major problems for signing When was the last time you checked a signature)
%       \item Interacting with communities
%     \end{itemize}

% \end{itemize}

\iffalse
	\myparagraph{Development of Systematic Incentive Theories}
	Developing systematic theories of incentives by conducting interviews with open-source software developers to gain insights into their motivations for signing, identify major challenges in the signing process, and inquire about their recent experiences in checking signatures, beyond platform-specific incentives. Additionally, engaging with software development communities to further understand the dynamics of incentive structures.
\fi

\myparagraph{Identifying Machine Learning (ML) Software and Pre-Trained Model Signing requirements}
Signing adoption rates in the Hugging Face registry are much lower than in all other studied registries.
We recommend research on signing practices in this context.
The issue might be an odd signing target --- commits rather than packages.
The challenge might be more fundamental, clarifying the nature of effective signatures for ML models and training regimes~\cite{davis2023reusing}.
%This is necessary since ML has different characteristics than traditional software. Therefore, understanding these differences is key to developing systems to sign them effectively.
% We also recommend implementing a similar solution like Sigstore for machine learning artifacts, taking note of the unique aspects of ML ecosystems.

\myparagraph{Apply the results}
As noted in~\cref{sec:discussion}, registry operators appear to have a strong influence on software signing quantity and quality.
Partnering with registry operators, researchers can apply our results to empirically validate them.

\section{Conclusion}

In this study, we assessed signing in four public software package registries (\registries).
We measured signature adoption (quantity and quality).
We found that, aside from \maven, the quantity of signatures in package registries is low.
Aside from \docker, the quality of signatures in package registries is low.
To explain these observations,
we proposed and evaluated an incentive-basd theory explaining maintainer's decisions to adopt signatures.
%we proposed and evaluated an incentive theory for signing adoption.
We used quasi-experiments to test four hypotheses.
We found that incentives do influence signing adoption, and some incentives are more influential than others.
Registry policies and startup costs seem to have the largest impact on signing adoption.
Cybersecurity events do not appear to have a significant impact on signing adoption.

We hope that our results will encourage the software engineering community to improve their software signing efforts, enhancing the overall security of software systems.
Our findings suggest specific incentives that could significantly improve software signing adoption rates.

\section{Data Availability}
% The artifacts for this paper are available at \placeholder{\url{https://github.com/PurdueDualityLab/signature-adoption}}.
The tools used to collect and analyze the data are available at \url{https://github.com/PurdueDualityLab/signature-adoption}.
This repository also contains the reported data in a relational database snapshot.

\ifCLASSOPTIONcompsoc
	\section*{Acknowledgments}
\else
	\section*{Acknowledgment}
\fi

This work was supported by Google, Cisco, and NSF award \#2229703.
S. Joshi helped with the statistical analysis.




\balance

\bibliographystyle{IEEEtran}
\bibliography{bib/master,bib/jamie}

\begin{thebibliography}{100}
\providecommand{\url}[1]{#1}
\csname url@samestyle\endcsname
\providecommand{\newblock}{\relax}
\providecommand{\bibinfo}[2]{#2}
\providecommand{\BIBentrySTDinterwordspacing}{\spaceskip=0pt\relax}
\providecommand{\BIBentryALTinterwordstretchfactor}{4}
\providecommand{\BIBentryALTinterwordspacing}{\spaceskip=\fontdimen2\font plus
\BIBentryALTinterwordstretchfactor\fontdimen3\font minus \fontdimen4\font\relax}
\providecommand{\BIBforeignlanguage}[2]{{%
\expandafter\ifx\csname l@#1\endcsname\relax
\typeout{** WARNING: IEEEtran.bst: No hyphenation pattern has been}%
\typeout{** loaded for the language `#1'. Using the pattern for}%
\typeout{** the default language instead.}%
\else
\language=\csname l@#1\endcsname
\fi
#2}}
\providecommand{\BIBdecl}{\relax}
\BIBdecl

\bibitem{gsa_18f_18f_nodate}
G.~18F, ``\BIBforeignlanguage{en-US}{{18F}: {Digital} service delivery {\textbar} {Open} source policy},'' \url{https://18f.gsa.gov/open-source-policy/}.

\bibitem{department_of_defense_open_2003}
D.~of~Defense, ``Open {Source} {Software} ({OSS}) in the {Department} of {Defense} ( {DoD}),'' May 2003.

\bibitem{synopsys_2023_2023}
Synopsys, ``\BIBforeignlanguage{en}{2023 {OSSRA} {Report}},'' Synopsys, Tech. Rep., 2023, \url{ https://www.synopsys.com/software-integrity/engage/ossra/rep-ossra-2023-pdf }.

\bibitem{okafor_sok}
C.~Okafor, T.~R. Schorlemmer, S.~Torres-Arias, and J.~C. Davis, ``Sok: Analysis of software supply chain security by establishing secure design properties,'' in \emph{The ACM Workshop on Software Supply Chain Offensive Research and Ecosystem Defenses (SCORED)}, 2022, p. 15–24.

\bibitem{solarwinds}
FireEye, ``Highly evasive attacker leverages solarwinds supply chain to compromise multiple global victims with sunburst backdoor,'' \url{ https://www.fireeye.com/blog/threat-research/2020/12/evasive-attacker-leverages-solarwinds-supply-chain-compromises-with-sunburst-backdoor.html }.

\bibitem{solarwinds-microsoft}
{Microsoft Security Response Center}, ``Customer guidance on recent nation-state cyber attacks,'' \url{ https://msrc-blog.microsoft.com/2020/12/13/customer-guidance-on-recent-nation-state-cyber-attacks/ }.

\bibitem{solarwinds-zdnet}
\BIBentryALTinterwordspacing
{Catalin Cimpanu}, ``Microsoft, fireeye confirm solarwinds supply chain attack,'' https://www.zdnet.com/article/microsoft-fireeye-confirm-solarwinds-supply-chain-attack/. [Online]. Available: \url{https://www.zdnet.com/article/microsoft-fireeye-confirm-solarwinds-supply-chain-attack/}
\BIBentrySTDinterwordspacing

\bibitem{sonatype_state_2022}
Sonatype, ``State of the {Software} {Supply} {Chain},'' Sonatype, Tech. Rep. 8th Annual, 2022, \url{ https://www.sonatype.com/state-of-the-software-supply-chain/open-source-supply-demand-security }.

\bibitem{zahan_openssf_2023}
N.~Zahan, P.~Kanakiya, B.~Hambleton, S.~Shohan, and L.~Williams, ``{OpenSSF} {Scorecard}: {On} the {Path} {Toward} {Ecosystem}-wide { Automated} {Security} {Metrics},'' Jan. 2023, \url{http://arxiv.org/abs/2208.03412}.

\bibitem{zahan2023softwarePractices}
N.~Zahan, S.~Shohan, D.~Harris, and L.~Williams, ``Do software security practices yield fewer vulnerabilities?'' in \emph{2023 IEEE/ACM 45th International Conf. on Software Engineering: Software Engineering in Practice (ICSE-SEIP)}.\hskip 1em plus 0.5em minus 0.4em\relax IEEE, 2023, pp. 292--303.

\bibitem{serebryany_oss-fuzz_2017}
\BIBentryALTinterwordspacing
K.~Serebryany, ``\BIBforeignlanguage{en}{Oss-fuzz --- {Google}'s continuous fuzzing service for open source software},'' 2017. [Online]. Available: \url{https://www.usenix.org/conference/usenixsecurity17/technical-sessions/presentation/serebryany}
\BIBentrySTDinterwordspacing

\bibitem{10.1145/3548606.3559373}
N.~Luo, T.~Antonopoulos, W.~R. Harris, R.~Piskac, E.~Tromer, and X.~Wang, ``Proving unsat in zero knowledge,'' in \emph{Proceedings of the ACM SIGSAC Conf. on Computer and Communications Security (CCS)}, 2022, p. 2203–2217.

\bibitem{balliu2023challenges}
M.~Balliu, B.~Baudry, S.~Bobadilla, M.~Ekstedt, M.~Monperrus, J.~Ron, A.~Sharma, G.~Skoglund, C.~Soto-Valero, and M.~Wittlinger, ``Challenges of producing software bill of materials for java,'' \emph{arXiv preprint arXiv:2303.11102}, 2023.

\bibitem{zahan2023software}
N.~Zahan, E.~Lin, M.~Tamanna, W.~Enck, and L.~Williams, ``Software bills of materials are required. are we there yet?'' \emph{IEEE Security \& Privacy}, vol.~21, no.~2, pp. 82--88, 2023.

\bibitem{winters_software_2020}
T.~Winters, T.~Manshreck, and H.~Wright, \emph{Software engineering at {Google}: lessons learned from programming over time}, first edition~ed.\hskip 1em plus 0.5em minus 0.4em\relax Beijing [China]: O'Reilly Media, 2020.

\bibitem{kuppusamy_diplomat}
T.~K. Kuppusamy, S.~Torres-Arias, V.~Diaz, and J.~Cappos, ``Diplomat: Using delegations to protect community repositories,'' in \emph{USENIX Symposium on Networked Systems Design and Implementation (NSDI)}, Santa Clara, CA, 2016, pp. 567--581.

\bibitem{woodruff_pgp_2023}
W.~Woodruff, ``\BIBforeignlanguage{en}{{PGP} signatures on {PyPI}: worse than useless},'' May 2023, \url{ https://blog.yossarian.net/2023/05/21/PGP-signatures-on-PyPI-worse-than-useless }.

\bibitem{felderer_contemporary_2020}
M.~Felderer and G.~H. Travassos, Eds., \emph{\BIBforeignlanguage{eng}{Contemporary empirical methods in software engineering}}.\hskip 1em plus 0.5em minus 0.4em\relax Cham: Springer, 2020.

\bibitem{Jiang2023PTMReuse}
W.~Jiang, N.~Synovic, M.~Hyatt, T.~R. Schorlemmer, R.~Sethi, Y.-H. Lu, G.~K. Thiruvathukal, and J.~C. Davis, ``An empirical study of pre-trained model reuse in the hugging face deep learning model registry,'' in \emph{{IEEE}/{ACM} 45th {International} {Conf.} on {Software} {Engineering} (ICSE'23)}, 2023.

\bibitem{sourceforge_compare_nodate}
SourceForge, ``Compare, {Download} \& {Develop} {Open} {Source} \& {Business} { Software} - {SourceForge},'' \url{https://sourceforge.net/}.

\bibitem{gitlab_devsecops_nodate}
GitLab, ``\BIBforeignlanguage{en-us}{The {DevSecOps} {Platform}},'' \url{https://about.gitlab.com/}.

\bibitem{GitHubPackages}
``{GitHub Packages},'' \url{https://github.com/features/packages}, accessed: 2023-12-06.

\bibitem{jiang2022PTMSupplyChain}
W.~Jiang, N.~Synovic, R.~Sethi, A.~Indarapu, M.~Hyatt, T.~R. Schorlemmer \emph{et~al.}, ``An empirical study of artifacts and security risks in the pre-trained model supply chain,'' in \emph{The ACM Workshop on Software Supply Chain Offensive Research and Ecosystem Defenses (SCORED)}, 2022, p. 105–114.

\bibitem{zimmermann2019small}
M.~Zimmermann, C.-A. Staicu, C.~Tenny, and M.~Pradel, ``Small world with high risks: A study of security threats in the npm ecosystem,'' in \emph{USENIX Security}, 2019, pp. 995--1010.

\bibitem{enck2022top}
W.~Enck and L.~Williams, ``Top 5 challenges in software supply chain security: Observations from 30 industry and government organizations,'' \emph{IEEE Sec. \& Privacy}, vol.~20, no.~2, pp. 96--100, 2022.

\bibitem{okafor_sok_2022}
C.~Okafor, T.~R. Schorlemmer, S.~Torres-Arias, and J.~C. Davis, ``\BIBforeignlanguage{en}{{SoK}: {Analysis} of {Software} {Supply} {Chain} {Security} by { Establishing} {Secure} {Design} {Properties}},'' in \emph{\BIBforeignlanguage{en}{{ACM} {Workshop} on {Software} {Supply} {Chain} {Offens.} {Res.} and {Ecosys.} {Defenses}}}, 2022, pp. 15--24.

\bibitem{kumar2017security}
D.~Kumar, Z.~Ma, Z.~Durumeric, A.~Mirian, J.~Mason, J.~A. Halderman, and M.~Bailey, ``Security challenges in an increasingly tangled web,'' in \emph{Proceedings of the 26th International Conf. on World Wide Web}, 2017, pp. 677--684.

\bibitem{Wang_Chen_Huang_Shi_Xu_Peng_Wu_Liu_2020}
\BIBentryALTinterwordspacing
Y.~Wang, B.~Chen, K.~Huang, B.~Shi, C.~Xu, X.~Peng, Y.~Wu, and Y.~Liu, ``\BIBforeignlanguage{en}{An empirical study of usages, updates and risks of third-party libraries in java projects},'' in \emph{\BIBforeignlanguage{en}{2020 IEEE International Conf. on Software Maintenance and Evolution (ICSME)}}.\hskip 1em plus 0.5em minus 0.4em\relax Adelaide, Australia: IEEE, Sep. 2020, p. 35–45. [Online]. Available: \url{https://ieeexplore.ieee.org/document/9240619/}
\BIBentrySTDinterwordspacing

\bibitem{de2008empirical}
C.~R. de~Souza and D.~F. Redmiles, ``An empirical study of software developers' management of dependencies and changes,'' in \emph{International Conf. on Software Engineering (ICSE)}, 2008, pp. 241--250.

\bibitem{pashchenko2020qualitative}
I.~Pashchenko, D.-L. Vu, and F.~Massacci, ``A qualitative study of dependency management and its security implications,'' in \emph{Proceedings of the 2020 ACM SIGSAC conference on computer and communications security}, 2020, pp. 1513--1531.

\bibitem{jadhav2009evaluating}
A.~S. Jadhav and R.~M. Sonar, ``Evaluating and selecting software packages: A review,'' \emph{Information and software technology}, vol.~51, no.~3, pp. 555--563, 2009.

\bibitem{jadhav2011framework}
------, ``Framework for evaluation and selection of the software packages: A hybrid knowledge based system approach,'' \emph{Journal of Systems and Software}, vol.~84, no.~8, pp. 1394--1407, 2011.

\bibitem{decan2019empirical}
A.~Decan, T.~Mens, and P.~Grosjean, ``An empirical comparison of dependency network evolution in seven software packaging ecosystems,'' \emph{Empirical Software Engineering}, vol.~24, pp. 381--416, 2019.

\bibitem{ghofrani2022trust}
J.~Ghofrani, P.~Heravi, K.~A. Babaei, and M.~D. Soorati, ``Trust challenges in reusing open source software: An interview-based initial study,'' in \emph{Proceedings of the 26th ACM International Systems and Software Product Line Conf.-Volume B}, 2022, pp. 110--116.

\bibitem{vu_lastpymile_2021}
D.-L. Vu, F.~Massacci, I.~Pashchenko, H.~Plate, and A.~Sabetta, ``{LastPyMile}: identifying the discrepancy between sources and packages,'' in \emph{The {ACM} {Joint} {Meeting} on {European} { Software} {Engineering} {Conf.} and {Symposium} on the { Foundations} of {Software} {Engineering} (ESEC/FSE)}, 2021, pp. 780--792.

\bibitem{intoto}
S.~Torres-Arias, H.~Afzali, T.~K. Kuppusamy, R.~Curtmola, and J.~Cappos, ``in-toto: Providing farm-to-table guarantees for bits and bytes,'' in \emph{USENIX Security Symposium}, 2019, pp. 1393--1410.

\bibitem{lamb_reproducible_2022}
C.~Lamb and S.~Zacchiroli, ``Reproducible {Builds}: {Increasing} the {Integrity} of {Software} { Supply} {Chains},'' \emph{IEEE Software}, vol.~39, no.~2, pp. 62--70, Mar. 2022, conf. Name: IEEE Software.

\bibitem{hejderup_use_2022}
J.~Hejderup, ``\BIBforeignlanguage{en}{On the {Use} of {Tests} for {Software} {Supply} {Chain} {Threats}},'' in \emph{\BIBforeignlanguage{en}{{ACM} {Workshop} on {Software} {Supply} {Chain} {Offensive} {Research} and {Ecosystem} {Defenses} (SCORED)}}, 2022, pp. 47--49.

\bibitem{benedetti_automatic_2022}
G.~Benedetti, ``\BIBforeignlanguage{en}{Automatic {Security} {Assessment} of {GitHub} {Actions} {Workflows}},'' \emph{\BIBforeignlanguage{en}{Los Angeles}}, 2022, \url{https://dl.acm.org/doi/abs/10.1145/3560835.3564554}.

\bibitem{noauthor_minimum_nodate}
\BIBentryALTinterwordspacing
``The {Minimum} {Elements} {For} a {Software} {Bill} of {Materials} ( {SBOM}) {\textbar} {National} {Telecommunications} and {Information} {Administration}.'' [Online]. Available: \url{https://www.ntia.gov/report/2021/minimum-elements-software-bill-materials-sbom}
\BIBentrySTDinterwordspacing

\bibitem{ohm_towards_2020}
\BIBentryALTinterwordspacing
M.~Ohm, A.~Sykosch, and M.~Meier, ``Towards detection of software supply chain attacks by forensic artifacts,'' in \emph{Proceedings of the 15th {International} {Conf.} on { Availability}, {Reliability} and {Security}}, ser. {ARES} '20.\hskip 1em plus 0.5em minus 0.4em\relax New York, NY, USA: Association for Computing Machinery, Aug. 2020, pp. 1--6. [Online]. Available: \url{https://doi.org/10.1145/3407023.3409183}
\BIBentrySTDinterwordspacing

\bibitem{shirey2007internet}
\BIBentryALTinterwordspacing
R.~Shirey, ``Internet security glossary, version 2,'' \url{https://www.rfc-editor.org/rfc/rfc4949.html}, Network Working Group, RFC 4949, August 2007, fYI: 36. Obsoletes: 2828. [Online]. Available: \url{https://www.rfc-editor.org/rfc/rfc4949.html}
\BIBentrySTDinterwordspacing

\bibitem{security_technical_advisory_group_software_2021}
\BIBentryALTinterwordspacing
S.~T.~A. Group, ``\BIBforeignlanguage{en}{Software {Supply} {Chain} {Best} {Practices}},'' Cloud Native Computing Foundation, Tech. Rep., May 2021. [Online]. Available: \url{https://project.linuxfoundation.org/hubfs/CNCF_SSCP_v1.pdf}
\BIBentrySTDinterwordspacing

\bibitem{slsa}
\BIBentryALTinterwordspacing
``\BIBforeignlanguage{en}{Supply-chain {Levels} for {Software} {Artifacts}}.'' [Online]. Available: \url{https://slsa.dev/}
\BIBentrySTDinterwordspacing

\bibitem{nist-2018-codesigning}
\BIBentryALTinterwordspacing
D.~Cooper, A.~Regenscheid, M.~Souppaya, C.~Bean, M.~Boyle, D.~Cooley, and M.~Jenkins, ``Security considerations for code signing,'' National Institute of Standards and Technology, Ft. George G. Meade, Maryland, NIST Cybersecurity White Paper, January 2018. [Online]. Available: \url{https://nvlpubs.nist.gov/nistpubs/CSWP/NIST.CSWP.01262018.pdf}
\BIBentrySTDinterwordspacing

\bibitem{cisa-2021-supplychain}
\BIBentryALTinterwordspacing
{Cybersecurity and Infrastructure Security Agency}, ``Defending against software supply chain attacks,'' Cybersecurity and Infrastructure Security Agency, Technical Report, April 2021. [Online]. Available: \url{https://www.cisa.gov/sites/default/files/publications/defending_against_software_supply_chain_attacks_508_1.pdf}
\BIBentrySTDinterwordspacing

\bibitem{gnu_privacy_guard}
\BIBentryALTinterwordspacing
T.~P. o. t.~G. Project, ``\BIBforeignlanguage{en}{The {GNU} {Privacy} {Guard}},'' Apr. 2023, publisher: The GnuPG Project. [Online]. Available: \url{https://gnupg.org/}
\BIBentrySTDinterwordspacing

\bibitem{gopalakrishna2022if}
N.~K. Gopalakrishna, D.~Anandayuvaraj, A.~Detti, F.~L. Bland \emph{et~al.}, ``"if security is required": Engineering and security practices for machine learning-based iot devices,'' in \emph{Internat'l. Workshop on Software Eng. Res. and Practice for the IoT}, 2022, pp. 1--8.

\bibitem{meng2018secure}
N.~Meng, S.~Nagy, D.~Yao, W.~Zhuang, and G.~A. Argoty, ``Secure coding practices in java: Challenges and vulnerabilities,'' in \emph{International Conf. on Software Engineering}, 2018, pp. 372--383.

\bibitem{chen2019reliable}
M.~Chen, F.~Fischer, N.~Meng, X.~Wang, and J.~Grossklags, ``How reliable is the crowdsourced knowledge of security implementation?'' in \emph{2019 IEEE/ACM 41st International Conf. on Software Engineering (ICSE)}.\hskip 1em plus 0.5em minus 0.4em\relax IEEE, 2019, pp. 536--547.

\bibitem{whitten_why_1999}
A.~Whitten and J.~D. Tygar, ``Why {Johnny} can't encrypt: a usability evaluation of {PGP} 5.0,'' in \emph{{USENIX} {Security} {Symposium}}, 1999, p.~14.

\bibitem{sheng_why_2016}
S.~Ruoti, J.~Andersen, D.~Zappala, and K.~Seamons, ``Why johnny still, still can't encrypt: Evaluating the usability of a modern pgp client,'' \emph{arXiv}, 2016, \url{https://arxiv.org/abs/1510.08555}.

\bibitem{Gillian_OpenITP}
{Gillian Andrews}, ``{Usability Report: Proposed Mailpile Features.}'' {OpenITP}, {December} {2014}, \url{ https://openitp.org/field-notes/user-tests-mailpile-features.html} { Accessed on 27/06/2023}.

\bibitem{braz_security_2006}
C.~Braz and J.-M. Robert, ``Security and usability: the case of the user authentication methods,'' in \emph{The 18th {Conf.} on l'{Interaction} {Homme} -{Machine} (IHM)}, 2006, pp. 199--203, \url{https://dl.acm.org/doi/10.1145/1132736.1132768}.

\bibitem{ruoti_confused_2013}
S.~Ruoti, N.~Kim, B.~Burgon, T.~van~der Horst, and K.~Seamons, ``Confused {Johnny}: when automatic encryption leads to confusion and mistakes,'' in \emph{Proceedings of the {Ninth} {Symposium} on {Usable} {Privacy} and {Security} (SOUPS)}.\hskip 1em plus 0.5em minus 0.4em\relax ACM, 2013, pp. 1--12.

\bibitem{reuter_secure_2020}
A.~Reuter, K.~Boudaoud, M.~Winckler, A.~Abdelmaksoud, and W.~Lemrazzeq, ``\BIBforeignlanguage{en}{Secure {Email} - {A} {Usability} {Study}},'' in \emph{\BIBforeignlanguage{en}{Financial {Cryptography} and {Data} {Security}}}, ser. Lecture {Notes} in {Computer} {Science}, M.~Bernhard, A.~Bracciali, L.~J. Camp, S.~Matsuo, A.~Maurushat, P.~B. Rønne, and M.~Sala, Eds.\hskip 1em plus 0.5em minus 0.4em\relax Cham: Springer International Publishing, 2020, pp. 36--46.

\bibitem{fahl_helping_2012}
S.~Fahl, M.~Harbach, T.~Muders, M.~Smith, and U.~Sander, ``Helping {Johnny} 2.0 to encrypt his {Facebook} conversations,'' in \emph{{Symposium} on {Usable} {Privacy} and {Security} (SOUPS)}, 2012, pp. 1--17.

\bibitem{ruoti_private_2015}
S.~Ruoti, J.~Andersen, T.~Hendershot, D.~Zappala, and K.~Seamons, ``Private {Webmail} 2.0: {Simple} and {Easy}-to-{Use} {Secure} {Email },'' Oct. 2015, \url{http://arxiv.org/abs/1510.08435}.

\bibitem{atwater_leading_2015}
E.~Atwater, C.~Bocovich, U.~Hengartner, E.~Lank, and I.~Goldberg, ``Leading {Johnny} to water: designing for usability and trust,'' in \emph{{USENIX} {Conf.} on {Usable} {Privacy} and {Security} (SOUPS)}, 2015.

\bibitem{latacora_pgp_2019}
Latacora, ``The {PGP} {Problem},'' Jul. 2019, \url{https://latacora.micro.blog/2019/07/16/the-pgp-problem.html}.

\bibitem{green_whats_2014}
M.~Green, ``\BIBforeignlanguage{en}{What's the matter with {PGP}?}'' Aug. 2014, \url{ https://blog.cryptographyengineering.com/2014/08/13/whats-matter-with-pgp/ }.

\bibitem{mouli-cloud}
R.~Chandramouli, M.~Iorga, and S.~Chokhani, ``Cryptographic key management issues and challenges in cloud services,'' \emph{Secure Cloud Computing}, pp. 1--30, 2013.

\bibitem{cooper_security_2018}
D.~Cooper, {Andrew Regenscheid}, {Murugiah Souppaya}, {Christopher Bean}, {Mike Boyle}, {Dorothy Cooley}, and { Michael Jenkins}, ``\BIBforeignlanguage{en}{Security {Considerations} for {Code} {Signing}},'' \emph{\BIBforeignlanguage{en}{NIST Cybersecurity White Paper}}, Jan. 2018, \url{ https://csrc.nist.rip/external/nvlpubs.nist.gov/nistpubs/CSWP/NIST.CSWP.01262018.pdf }.

\bibitem{newman_sigstore_2022}
Z.~Newman, J.~S. Meyers, and S.~Torres-Arias, ``Sigstore: {Software} {Signing} for {Everybody},'' in \emph{The {ACM} {SIGSAC} {Conf.} on {Computer} and {Communications} {Security} (CCS)}, 2022.

\bibitem{melara2015coniks}
M.~S. Melara, A.~Blankstein, J.~Bonneau, E.~W. Felten, and M.~J. Freedman, ``Coniks: Bringing key transparency to end users,'' in \emph{24th USENIX Security Symposium}, 2015, pp. 383--398.

\bibitem{malvai2023parakeet}
H.~Malvai, L.~Kokoris-Kogias, A.~Sonnino, E.~Ghosh, E.~Ozt{\"u}rk, K.~Lewi, and S.~Lawlor, ``Parakeet: Practical key transparency for end-to-end encrypted messaging,'' \emph{Cryptology ePrint Archive}, 2023.

\bibitem{heftrig2022poster}
E.~Heftrig, H.~Shulman, and M.~Waidner, ``Poster: The unintended consequences of algorithm agility in dnssec,'' in \emph{The SIGSAC Conf. on Computer and Communications Security (CCS)}, 2022.

\bibitem{signature-distribution-update-hotsec}
A.~Bellissimo, J.~Burgess, and K.~Fu, ``Secure software updates: Disappointments and new challenges,'' in \emph{USENIX Workshop on Hot Topics in Security (HotSec)}, 2006.

\bibitem{merrill_speranza_2023}
K.~Merrill, Z.~Newman, S.~Torres-Arias, and K.~Sollins, ``Speranza: {Usable}, privacy-friendly software signing,'' May 2023, \url{http://arxiv.org/abs/2305.06463}.

\bibitem{OpenBSD}
OpenBSD, ``signify: Securing openbsd from us to you,'' "\url{https://www.openbsd.org/papers/bsdcan-signify.html}".

\bibitem{brown_digital_nudges_2021}
\BIBentryALTinterwordspacing
D.~C. Brown, ``\BIBforeignlanguage{English}{Digital nudges for encouraging developer behaviors},'' Ph.D. dissertation, 2021, copyright - Database copyright ProQuest LLC; ProQuest does not claim copyright in the individual underlying works; Last updated - 2023-07-19. [Online]. Available: \url{https://www.proquest.com/dissertations-theses/digital-nudges-encouraging-developer-behaviors/docview/2566242524/se-2}
\BIBentrySTDinterwordspacing

\bibitem{deci_effects_1971}
E.~L. Deci, ``Effects of externally mediated rewards on intrinsic motivation,'' \emph{Journal of Personality and Social Psychology}, vol.~18, no.~1, pp. 105--115, 1971.

\bibitem{baddeley_behavioural_2017}
\BIBentryALTinterwordspacing
M.~Baddeley, \emph{{Behavioural Economics: A Very Short Introduction}}.\hskip 1em plus 0.5em minus 0.4em\relax Oxford University Press, 01 2017. [Online]. Available: \url{https://doi.org/10.1093/actrade/9780198754992.001.0001}
\BIBentrySTDinterwordspacing

\bibitem{kamenica_behavioral_economics_2012}
\BIBentryALTinterwordspacing
E.~Kamenica, ``Behavioral economics and psychology of incentives,'' \emph{Annual Review of Economics}, vol.~4, no.~1, pp. 427--452, 2012. [Online]. Available: \url{https://doi.org/10.1146/annurev-economics-080511-110909}
\BIBentrySTDinterwordspacing

\bibitem{gneezy_pay_2000}
\BIBentryALTinterwordspacing
U.~Gneezy and A.~Rustichini, ``Pay enough or don't pay at all,'' \emph{The Quarterly Journal of Economics}, vol. 115, no.~3, pp. 791--810, 2000. [Online]. Available: \url{http://www.jstor.org/stable/2586896}
\BIBentrySTDinterwordspacing

\bibitem{titmuss_gift_2018}
\BIBentryALTinterwordspacing
R.~M. Titmuss, \emph{The gift relationship (reissue): From human blood to social policy}, 1st~ed.\hskip 1em plus 0.5em minus 0.4em\relax Bristol University Press, 2018. [Online]. Available: \url{http://www.jstor.org/stable/j.ctv6zdcmh}
\BIBentrySTDinterwordspacing

\bibitem{gotterbarn1997software}
D.~Gotterbarn, K.~Miller, and S.~Rogerson, ``Software engineering code of ethics,'' \emph{Comm. of the ACM}, vol.~40, no.~11, 1997.

\bibitem{remler_research_2015}
D.~K. Remler and G.~G. Van~Ryzin, \emph{Research methods in practice: strategies for description and causation}, second edition~ed.\hskip 1em plus 0.5em minus 0.4em\relax Los Angeles: SAGE, 2015.

\bibitem{wohlin_experimentation_2012}
C.~Wohlin, \emph{Experimentation in software engineering}.\hskip 1em plus 0.5em minus 0.4em\relax New York: Springer, 2012.

\bibitem{wooldridge_introductory_2013}
J.~M. Wooldridge, \emph{Introductory econometrics: a modern approach}, 5th~ed.\hskip 1em plus 0.5em minus 0.4em\relax Mason, OH: South-Western Cengage Learning, 2013.

\bibitem{meyer_natural_1995}
B.~D. Meyer, ``Natural and {Quasi}-{Experiments} in {Economics},'' \emph{Journal of Business \& Economic Statistics}, vol.~13, no.~2, pp. 151--161, 1995, \url{https://www.jstor.org/stable/1392369}.

\bibitem{bogart_2021}
\BIBentryALTinterwordspacing
C.~Bogart, C.~K\"{a}stner, J.~Herbsleb, and F.~Thung, ``When and how to make breaking changes: Policies and practices in 18 open source software ecosystems,'' \emph{ACM Trans. Softw. Eng. Methodol.}, vol.~30, no.~4, jul 2021. [Online]. Available: \url{https://doi.org/10.1145/3447245}
\BIBentrySTDinterwordspacing

\bibitem{cass2023top}
\BIBentryALTinterwordspacing
S.~Cass, ``The top programming languages 2023,'' \url{https://spectrum.ieee.org/the-top-programming-languages-2023}, August 2023, accessed: 2023-12-07. [Online]. Available: \url{https://spectrum.ieee.org/the-top-programming-languages-2023}
\BIBentrySTDinterwordspacing

\bibitem{pahl2017cloud}
C.~Pahl, A.~Brogi, J.~Soldani, and P.~Jamshidi, ``Cloud container technologies: a state-of-the-art review,'' \emph{IEEE Transactions on Cloud Computing}, vol.~7, no.~3, pp. 677--692, 2017.

\bibitem{pypi_website}
{PyPI}, ``The python package index,'' 2024, \url{https://pypi.org/}.

\bibitem{stufft_pgp_2018}
\BIBentryALTinterwordspacing
D.~Stufft, ``\BIBforeignlanguage{en}{{PGP} signatures are not displayed {\textbar} {Issue} \#3356},'' Mar. 2018. [Online]. Available: \url{https://github.com/pypi/warehouse/issues/3356}
\BIBentrySTDinterwordspacing

\bibitem{stufft_removing_2023}
\BIBentryALTinterwordspacing
------, ``\BIBforeignlanguage{en}{Removing {PGP} from {PyPI} - {The} {Python} {Package} {Index}},'' May 2023. [Online]. Available: \url{https://blog.pypi.org/posts/2023-05-23-removing-pgp/}
\BIBentrySTDinterwordspacing

\bibitem{ecosystems_website}
{Open Collective}, ``Ecosyste.ms,'' 2024, \url{https://ecosyste.ms/}.

\bibitem{relations_pgp}
\BIBentryALTinterwordspacing
S.~D. Relations, ``\BIBforeignlanguage{en-us}{{PGP} vs. sigstore: {A} {Recap} of the {Match} at {Maven} {Central}},'' 2023. [Online]. Available: \url{https://blog.sonatype.com/pgp-vs.-sigstore-a-recap-of-the-match-at-maven-central}
\BIBentrySTDinterwordspacing

\bibitem{docker_dct}
\BIBentryALTinterwordspacing
Docker, ``\BIBforeignlanguage{en}{Content trust in {Docker}},'' 0200, \url{https://docs.docker.com/engine/security/trust/}. [Online]. Available: \url{https://docs.docker.com/engine/security/trust/}
\BIBentrySTDinterwordspacing

\bibitem{lorenc_cosign_2021}
\BIBentryALTinterwordspacing
D.~Lorenc, ``Cosign 1.0! - {Sigstore} {Blog},'' Jul. 2021, \url{https://blog.sigstore.dev/cosign-1-0-e82f006f7bc4/}. [Online]. Available: \url{https://blog.sigstore.dev/cosign-1-0-e82f006f7bc4/}
\BIBentrySTDinterwordspacing

\bibitem{hugging_face_website}
{Hugging Face}, ``Hugging face,'' 2024, \url{https://huggingface.co/}.

\bibitem{hugging_face_security}
\BIBentryALTinterwordspacing
{H}ugging {F}ace, ``Security,'' 2023. [Online]. Available: \url{https://huggingface.co/docs/hub/security}
\BIBentrySTDinterwordspacing

\bibitem{pypi_bigquery}
{PyPA}, ``Analyzing pypi package downloads,'' 2024, \url{https://packaging.python.org/en/latest/guides/analyzing-pypi-package-downloads/#public-dataset}.

\bibitem{notary}
\BIBentryALTinterwordspacing
``Notice,'' Oct. 2023, original-date: 2015-06-19T20:07:53Z. [Online]. Available: \url{https://github.com/notaryproject/notary}
\BIBentrySTDinterwordspacing

\bibitem{sonatype_pgp}
{Sonatype}, ``Working with pgp signatures,'' 2024, \url{https://central.sonatype.org/publish/requirements/gpg/#signing-a-file}.

\bibitem{NIST_RSA_SIZES}
\BIBentryALTinterwordspacing
H.~Ferraiolo and A.~Regenscheid, \emph{Cryptographic Algorithms and Key Sizes for Personal Identity Verification}, Sep. 2023. [Online]. Available: \url{http://dx.doi.org/10.6028/NIST.SP.800-78-5.ipd}
\BIBentrySTDinterwordspacing

\bibitem{hf_forum_post}
{Hugging Face}, ``Location of public gpg keys,'' 2024, \url{https://discuss.huggingface.co/t/location-of-public-gpg-keys/62915}.

\bibitem{anandayuvaraj2022reflecting}
D.~Anandayuvaraj and J.~C. Davis, ``Reflecting on recurring failures in iot development,'' in \emph{The 37th IEEE/ACM International Conf. on Automated Software Engineering}, 2022, pp. 1--5.

\bibitem{davis2023reusing}
J.~C. Davis, P.~Jajal, W.~Jiang, T.~R. Schorlemmer, N.~Synovic, and G.~K. Thiruvathukal, ``Reusing deep learning models: Challenges and directions in software engineering,'' in \emph{Proceedings of the IEEE John Vincent Atanasoff Symposium on Modern Computing}, 2023.

\end{thebibliography}

\newpage % The Meta-Review should at least start on a new column

% Use \appendices and not \appendix due to IEEEtran.cls quirks
\appendices % if not used earlier

\section{Meta-Review}

The following meta-review was prepared by the program committee for the 2024
IEEE Symposium on Security and Privacy (S\&P) as part of the review process as
detailed in the call for papers.

\subsection{Summary}
Code that can be authenticated to its source is a necessary prerequisite to a secure software supply chain.
This paper looks at package signing practices in PyPI, Maven, Hugging Face, and DockerHub, and evaluates the quality and quantity of the signatures found there.
The authors then compare the ecosystems to argue that only a mandate will lead to universal signing and ecosystems with easy-to-use tooling will lead to more voluntary signing.

\subsection{Scientific Contributions}
\begin{itemize}
	\item Provides a New Data Set For Public Use
	\item Addresses a Long-Known Issue
	\item Provides a Valuable Step Forward in an Established Field
\end{itemize}

\subsection{Reasons for Acceptance}
The paper assimilates significant data from multiple ecosystems that helps to shed light on signing practices.
The findings are interesting, showing a large amount of variance across ecosystems.
Likewise, rigid adherence to signing practices is also shown not to be a security panacea.
The data and corresponding insights struck the PC as a valuable contribution to aid understanding and direct future research.

\subsection{Noteworthy Concerns} % Exclude if your meta-review does not have noteworthy concerns
\begin{enumerate} % Enumerate environment is not necessary if there is only one
	\item While any empirical study will have limitations, the PC had concerns with the approach taken to classify public keys as valid. Best practices recommend that keys are rotated. The validity of a key seems most important at the time when a commit is performed, not years later. 
	\item The paper studies events that may have influenced signature adoption. It reports correlative, not causative, evidence for the influence of factors. Understanding causation requires further study.
	\item The paper excludes Hugging Face models that require assenting to a Terms of Service agreement. The PC wonders whether models with a ToS are more likely to be professionally developed and possibly more likely to be signed.
\end{enumerate}

\section{Response to the Meta-Review} % Optional

We take no issue with the meta-review.
We appreciate the holistic critique.

\end{document}